\documentclass[showpacs,preprintnumbers,superscriptaddress,preprint]{revtex4}
\usepackage[T1]{fontenc}     
\usepackage{amsfonts}
\usepackage{amsmath,amsbsy,amssymb,graphicx}               
\usepackage{times} 

\begin{document} 

\title{Nambu-Goldstone modes and the Josephson supercurrent 
in the bilayer quantum Hall system}  
\author{Yusuke~Hama}
\affiliation{Department of Physics, The University of Tokyo, Tokyo 113-0033,  Japan}
\affiliation{Theoretical Research Division, Nishina Center, RIKEN, Wako 351-0198, Japan}
\author{George~Tsitsishvili}
\affiliation{Department of Physics, Tbilisi State University, Tbilisi 0128, Georgia}
\author{Zyun~F.~Ezawa}
\affiliation{Advanced Meson Science Laborary, Nishina Center, RIKEN, Wako 351-0198, Japan}
\date{\today}

\date{\today} 
\preprint{RIKEN-QHP-81}  
\begin{abstract}%
An interlayer phase coherence develops spontaneously 
in the bilayer quantum Hall system at the filling factor $\nu =1$. 
On the other hand, the spin and pseudospin degrees of freedom are entangled coherently
in the canted antiferromagnetic phase of the bilayer quantum Hall system at
the filling factor $\nu =2$. 
There emerges a complex Nambu-Goldstone mode with a linear dispersion in the zero
tunneling-interaction limit  for both cases. 
Then its phase field provokes a Josephson supercurrent in each layer, 
which is dissipationless as in a superconductor. 
We study what kind of phase coherence the Nambu-Goldstone mode develops in association with
the Josephson supercurrent and its effect on the Hall resistance in the bilayer quantum Hall system
at $\nu=1,2$,
by employing the Grassmannian formalism.    
\end{abstract}

\maketitle

\section{Introduction}       
Quantum Hall (QH) effects are remarkable macroscopic quantum phenomena observed in the 2-dimensional
electron system\cite{kvKlitzing1,dcTsui1}. They are so special in condensed matter physics that they are deeply
connected with the fundamental principles of physics. Moreover, QH system provides us with an 
opportunity to enjoy the interplay between condensed matter physics and  particle and nuclear physics\cite{Ezawa:2008ae}.  

In particular, the physics of the bilayer quantum Hall (QH) system is enormously rich owing to
the intralayer and interlayer phase coherence controlled by the interplay
between the spin and the layer (pseudospin) degrees of freedom\cite{Ezawa:2008ae,S. Das Sarma1}. 
The interlayer phase coherence is an especially  intriguing phenomenon
in the bilayer QH system \cite{Ezawa:2008ae}, where it is enhanced
in the limit $\Delta_{\text{SAS}}\rightarrow 0$. 
For instance, at the filling factor $\nu =1$ there arises a 
unique phase, the spin-ferromagnet and pseudospin-ferromagnet phase, which
has  been well studied both theoretically and experimentally. One of the most
intriguing phenomena is the Josephson tunneling between the two layers
predicted in Refs.\cite{Ezawa:1992,Wen}, whose first experimental indication
was obtained in Ref.\cite{Spielman1}. Other examples are the anomalous
behavior of the Hall resistance reported in counterflow experiments\cite{Kellog1,Tutuc} 
and in drag experiments\cite{Kellog2}. They are triggered by
the Josephson supercurrent within each layer\cite{Ezawa:2007nj}. Quite
recently, careful experiments \cite{Tiemann} were performed to explore the
condition for the tunneling current to be dissipationless. These phenomena
are produced by the pseudospins at $\nu =1$, where the Nambu-Goldstone (NG) mode
describes a pseudospin wave.

On the other hand, at $\nu =2$ the bilayer QH system has three phases, the
spin-ferromagnet and pseudospin-singlet phase (abridged as the spin phase), 
the spin-singlet and pseudospin ferromagnet phase (abridged as the pseudospin phase) 
and a canted antiferromagnetic phase\cite{S. Das Sarma2,Pellegrini1,khrapai1,Sawada4} (abridged as the CAF phase),
depending on the relative strength between the Zeeman energy 
$\Delta_{\text{Z}}$ and the interlayer tunneling energy $\Delta_{\text{SAS}}$. 
The pattern of the symmetry breaking is 
SU(4)$\rightarrow $U(1)$\otimes $SU(2)$\otimes $SU(2), 
associated with which there appear four complex NG
modes\cite{Hasebe}. We have recently analyzed the full details of these
NG modes in each phase\cite{yhama}. The CAF phase is most
interesting, where one of the NG modes becomes gapless and has a
linear dispersion relation\cite{yhama} as the tunneling interaction vanishes
($\Delta_{\text{SAS}}\rightarrow 0$). It is an urgent and intriguing
problem what kind of phase coherence this NG mode develops.

In this paper, we investigate the interlayer phase coherence, the associated NG modes, 
its effective Hamiltonian, 
the Josephson supercurrent provoked by these NG modes and its effect to the Hall resistance 
in the bilayer QH system at $\nu=1,2$, by employing the
Grassmannian formalism\cite{Hasebe}.

The basic field is the Grassmannian field consisting of complex projective ($\text{CP}^{3}$) fields.  
We introduce $n$ $\text{CP}^{3}$
fields to analyze the $\nu=n$ bilayer QH system.  
The CP$^{3}$ field emerges when composite bosons undergo
Bose-Einstein condensation\cite{Ezawa:2008ae}. 
We first make a perturbative analysis of the NG modes and 
reproduce the same results as obtained in \cite{yhama}. 
We next analyze the nonperturbative phase 
coherent phenomena developed by the NG mode having linear dispersion,
 where the phase field $\vartheta (\mathbf{x})$ is
essentially classical and may become very large, which is necessary to analyze the associated Josephson
supercurrent. We show that it is the entangled spin-pseudospin phase coherence in the
CAF phase. The Grassmannian formalism provides us
with a clear physical picture of the spin-pseudospin phase coherence in the CAF phase, 
and, furthermore, enables us to describe nonperturbative  phase-coherent phenomena 
uniformly in the bilayer QH system.

We then show that the Josephson supercurrent flows within the layer when there is inhomogeneity in $\vartheta (\mathbf{x})$. 
A related topic has been investigated in \cite{yhama2}.
The supercurrent in the CAF phase leads to the same formula\cite{Ezawa:2007nj} 
for the anomalous Hall resistivity for the counterflow and
drag geometries as the one at $\nu =1$. What is remarkable is that the total
current flowing in the CAF phase is a Josephson supercurrent  carrying solely 
spins in the counterflow geometry. We also remark that the supercurrent
flows both in the balanced and imbalanced systems at $\nu =1$ but only in
imbalanced systems at $\nu =2$.

\section{The SU(4) effective Hamiltonian}     

Electrons in a plane perform cyclotron motion under perpendicular magnetic
field $B_{\perp }$ and create Landau levels. The number of flux quanta
passing through the system is $N_{\Phi }\equiv B_{\perp }S/\Phi _{\text{D}}$, 
where $S$ is the area of the system and $\Phi _{\text{D}}=2\pi \hbar /e$
is the flux quantum. There are $N_{\Phi }$ Landau sites per one Landau
level, each of which is associated with one flux quantum and occupies an
area $S/N_{\Phi }=2\pi \ell _{B}^{2}$, with the magnetic length 
$\ell _{B}=\sqrt{\hbar /eB_{\perp }}$.

In the bilayer system an electron has two types of index, the spin index 
$(\uparrow ,\downarrow )$ and the layer index $(\text{f},\text{b})$. They can
be incorporated in four types of isospin index, 
$\alpha =$ f$\uparrow $,f$\downarrow $,b$\uparrow $,b$\downarrow $. 
One Landau site may contain four
electrons. The filling factor is $\nu =N/N_{\Phi }$ with $N$ the total
number of electrons.

We explore the physics of electrons confined to the lowest Landau level (LLL),
where the electron position is specified solely by the guiding center 
$\boldsymbol{X}=(X,Y)$, whose $X$ and $Y$ components are noncommutative, 
\begin{equation}
\lbrack X,Y]=-i\ell _{B}^{2}.  \label{AlgebGC} 
\end{equation}
The equations of motion follow from this noncommutative relation rather than the kinetic term for electrons confined within the LLL.
In order to derive the effective Hamiltonian, it is convenient to represent the noncommutative relation with the use of the Fock states,
\begin{equation}
|n\rangle =\frac{1}{\sqrt{n!}}(b^{\dag })^{n}|0\rangle ,\quad n=0,1,2,\ldots
,\quad b|0\rangle =0,  \label{LandaSite}
\end{equation}
where $b$ and $b^{\dag }$ are the ladder operators, 
\begin{equation}
b=\frac{1}{\sqrt{2}\ell _{B}}(X-iY),\qquad b^{\dag }=\frac{1}{\sqrt{2}\ell_{B}}(X+iY),  \label{WeylXb}
\end{equation}
obeying $[b,b^{\dag }]=1$.
Although the Fock states correspond to the Landau sites in the symmetric gauge,
the resulting effective Hamiltonian is independent of the representation
we have chosen.

We expand the electron field operator by a complete set of one-body wave
functions $\varphi _{n}(\boldsymbol{x})=\langle \boldsymbol{x}|n\rangle $ in the LLL, 
\begin{equation}
\psi _{\alpha }(\boldsymbol{x})\equiv \sum_{n=1}^{N_{\Phi }}c_{\alpha}(n)\varphi _{n}(\boldsymbol{x}),  \label{electronfield}
\end{equation}
where $c_{\alpha }(n)$ is the annihilation operator at the Landau site 
$|n\rangle $ with $\alpha =$ f$\uparrow $,f$\downarrow $,b$\uparrow $,b$\downarrow $. 
The operators $c_{\alpha }(m),c_{\beta }^{\dagger }(n)$
satisfy the standard anticommutation relations,
\begin{align}
\{c_{\alpha }(m),c_{\beta }^{\dagger }(n)\}& =\delta _{mn}\delta _{\alpha\beta },  \quad
\{c_{\alpha }(m),c_{\beta }(n)\} =\{c_{\alpha }^{\dagger }(m),c_{\beta}^{\dagger }(n)\}=0.  
\end{align}
The electron field $\psi _{\alpha }(\boldsymbol{x})$ has four components, and
the bilayer system possesses the underlying algebra SU$(4)$, having the
subalgebra $\text{SU}_{\text{spin}}(2)\times \text{SU}_{\text{ppin}}(2)$. 
We denote the three generators of the $\text{SU}_{\text{spin}}(2)$ by $\tau
_{a}^{\text{spin}}$, and those of $\text{SU}_{\text{ppin}}(2)$ by $\tau
_{a}^{\text{ppin}}$. There  remain nine generators 
$\tau _{a}^{\text{spin}}\tau _{b}^{\text{ppin}}$,
whose explicit form is given in Appendix A.

All the physical operators required for the description of the system are
constructed as  bilinear combinations of $\psi (\boldsymbol{x})$ and 
$\psi^{\dagger }(\boldsymbol{x})$. They are 16 density operators  
\begin{align}
\rho (\boldsymbol{x})& =\psi ^{\dagger }(\boldsymbol{x})\psi (\boldsymbol{x}),  \quad
S_{a}(\boldsymbol{x}) =\frac{1}{2}\psi ^{\dagger }(\boldsymbol{x})\tau _{a}^{\text{spin}}
\psi (\boldsymbol{x}),  \notag \\
P_{a}(\boldsymbol{x})& =\frac{1}{2}\psi ^{\dagger }(\boldsymbol{x})\tau _{a}^{\text{ppin}}
\psi (\boldsymbol{x}),  \quad
R_{ab}(\boldsymbol{x}) =\frac{1}{2}\psi ^{\dagger }(\boldsymbol{x})\tau _{a}^{\text{spin}}
\tau _{b}^{\text{ppin}}\psi (\boldsymbol{x}),  \label{su4isospin1}
\end{align}
where $S_{a}$ describes the total spin and $2P_{z}$ measures the
electron-density difference between the two layers. The operator $R_{ab}$
transforms as a spin under $\text{SU}_{\text{spin}}(2)$ and as a pseudospin
under $\text{SU}_{\text{ppin}}(2)$.  
 
The kinetic Hamiltonian is quenched, since the kinetic energy is common to
all states in the LLL. The Coulomb interaction is decomposed into the
SU(4)-invariant and SU(4)-noninvariant terms 
\begin{align}
H_{\text{C}}^{+}& =\frac{1}{2}\int d^{2}xd^{2}yV^{+}(\boldsymbol{x}-\boldsymbol{y})
\rho (\boldsymbol{x})\rho (\boldsymbol{y}),  \label{coulomb1} \\
H_{\text{C}}^{-}& =2\int d^{2}xd^{2}yV^{-}(\boldsymbol{x}-\boldsymbol{y})P_{z}(\boldsymbol{x})
P_{z}(\boldsymbol{y}),  \label{coulomb2}
\end{align}
where 
\begin{equation}
V^{\pm }(\boldsymbol{x})=\frac{e^{2}}{8\pi \epsilon }\left( \frac{1}{|\boldsymbol{x}|}
\pm \frac{1}{\sqrt{|\boldsymbol{x}|^{2}+d^{2}}}\right),
\end{equation}
with  layer separation $d$. The tunneling and bias terms are summarized
into the pseudo-Zeeman term. Combining the Zeeman and pseudo-Zeeman terms 
we have 
\begin{equation}
H_{\text{ZpZ}}=-\int d^{2}x(\Delta _{\text{Z}}S_{z}+\Delta _{\text{SAS}}P_{x}+\Delta _{\text{bias}}P_{z}),
\end{equation}
with the Zeeman gap $\Delta _{\text{Z}}$, the tunneling gap $\Delta _{\text{SAS}}$, 
and the bias voltage $\Delta _{\text{bias}}=eV_{\text{bias}}$.

The total Hamiltonian is 
\begin{equation}
H=H_{\text{C}}^{+}+H_{\text{C}}^{-}+H_{\text{ZpZ}}. 
\end{equation}
Note that the SU(4)-noninvariant terms
vanish in the limit $d$, $\Delta _{\text{Z}}$, $\Delta _{\text{SAS}}$, $\Delta _{\text{bias}}\rightarrow 0$.

We project the density operators (\ref{su4isospin1}) to the LLL by
substituting the field operator (\ref{electronfield}) into them. A typical
density operator reads
\begin{equation}
R_{ab}(\boldsymbol{p})=e^{-\ell _{B}^{2}\boldsymbol{p}^{2}/4}\hat{R}_{ab}(\boldsymbol{p}),
\end{equation}
in  momentum space, with
\begin{equation}
\hat{R}_{ab}(\boldsymbol{p})=\frac{1}{4\pi }\sum_{mn}
\langle n|e^{-i\boldsymbol{pX}}|m\rangle c^{\dag }(n)\tau _{a}^{\text{spin}}
\tau _{b}^{\text{ppin}}c(m),  \label{BareDensiR}
\end{equation}
where $c(m)$ is the $4$-component vector made of the operators $c_{\alpha}(m)$.

What are observed experimentally are the classical densities, which are
expectation values such as $\hat{\rho}^{\text{cl}}(\boldsymbol{p})=\langle 
\mathfrak{S}|\hat{\rho}(\boldsymbol{p})|\mathfrak{S}\rangle $, 
where $|\mathfrak{S}\rangle $ represents a generic state in the LLL.   
The Coulomb Hamiltonian governing the classical densities are given by\cite{Ezawa:2003sr}: 
\begin{align}
H^{\text{eff}}& =\pi \int d^{2}pV_{D}^{+}(\boldsymbol{p})\hat{\rho}^{\text{cl}}
(-\boldsymbol{p})\hat{\rho}^{\text{cl}}(\boldsymbol{p})  
 +4\pi \int d^{2}pV_{D}^{-}(\boldsymbol{p})\hat{P}_{z}^{\text{cl}}(-\boldsymbol{p})
\hat{P}_{z}^{\text{cl}}(\boldsymbol{p})  \notag \\
& -\frac{\pi }{2}\int d^{2}pV_{X}^{d}(\boldsymbol{p})[\hat{S}_{a}^{\text{cl}}
(-\boldsymbol{p})\hat{S}_{a}^{\text{cl}}(\boldsymbol{p})+\hat{P}_{a}^{\text{cl}}
(-\boldsymbol{p})\hat{P}_{a}^{\text{cl}}(\boldsymbol{p})  
 +\hat{R}_{ab}^{\text{cl}}(-\boldsymbol{p})\hat{R}_{ab}^{\text{cl}}(\boldsymbol{p})]\notag\\
&-\pi\int d^{2}pV_{X}^{-}(\boldsymbol{p})[\hat{S}_{a}^{\text{cl}}(-\boldsymbol{p})
\hat{S}_{a}^{\text{cl}}(\boldsymbol{p})  
 +\hat{P}_{z}^{\text{cl}}(-\boldsymbol{p})\hat{P}_{z}^{\text{cl}}(\boldsymbol{p})+
\hat{R}_{az}^{\text{cl}}(-\boldsymbol{p})\hat{R}_{az}^{\text{cl}}(\boldsymbol{p})] 
\notag \\
& -\frac{\pi }{8}\int d^{2}pV_{X}(\boldsymbol{p})\hat{\rho}^{\text{cl}}(-\boldsymbol{p})
\hat{\rho}^{\text{cl}}(\boldsymbol{p}),
\end{align}
where $V_{D}$ and $V_{X}$ are the direct and exchange Coulomb potentials,
respectively, 
\begin{align}
V_{D}(\boldsymbol{p})& =\frac{e^{2}}{4\pi \epsilon |\boldsymbol{p}|}e^{-\ell _{B}^{2}
\boldsymbol{p}^{2}/2},  \quad
V_{X}(\boldsymbol{p}) =\frac{\sqrt{2\pi }e^{2}\ell_{B}}{4\pi \epsilon }I_{0}
(\ell_{B}^{2}\boldsymbol{p}^{2}/4)e^{-\ell _{B}^{2}\boldsymbol{p}^{2}/4},
\end{align}
with  $V_{X}=V_{X}^{+}+V_{X}^{-},\quad V_{X}^{d}=V_{X}^{+}-V_{X}^{-}$, and
\begin{align}
V_{D}^{\pm }(\boldsymbol{p})& =\frac{e^{2}}{8\pi \epsilon |\boldsymbol{p}|}
\left( 1\pm e^{-|\boldsymbol{p}|d}\right) e^{-\ell _{B}^{2}\boldsymbol{p}^{2}/2},  
\notag\\
V_{X}^{\pm }(\boldsymbol{p})& =\frac{\sqrt{2\pi }e^{2}\ell _{B}}{8\pi\epsilon }
I_{0}(\ell _{B}^{2}\boldsymbol{p}^{2}/4)e^{-\ell _{B}^{2}
\boldsymbol{p}^{2}/4} 
 \pm \frac{e^{2}\ell _{B}^{2}}{4\pi \epsilon }\int_{0}^{\infty }dke^{-\frac{1}{2}
\ell _{B}^{2}k^{2}-kd}J_{0}(\ell _{B}^{2}|\boldsymbol{p}|k).
\end{align}%
Here, $I_{0}(x)$ is the modified Bessel function, and $J_{0}(x)$ is the
Bessel function of the first kind.

Since the exchange interaction $V^{\pm }(\boldsymbol{p})$ is short ranged, it is
a good approximation to make the derivative expansion, or, equivalently, the
momentum expansion. 
We may set $\hat{\rho}^{\text{cl}}(\boldsymbol{p})=\rho _{0}$, 
$\hat{S}_{a}^{\text{cl}}(\boldsymbol{p})=\rho _{\Phi }\mathcal{S}_{a}(\boldsymbol{p})$,
$\hat{P}_{a}^{\text{cl}}(\boldsymbol{p})=\rho _{\Phi }\mathcal{P}_{a}(\boldsymbol{p})$, 
and $\hat{R}_{ab}^{\text{cl}}(\boldsymbol{p})=\rho _{\Phi }\mathcal{R}_{ab}(\boldsymbol{p})$ 
for the study of NG modes. Taking the nontrivial lowest-order
terms in the derivative expansion, we obtain the SU(4) effective Hamiltonian
density 
\begin{align} 
{\mathcal{H}}^{\text{eff}}& =J_{s}^{d}\left( \sum (\partial _{k}
\mathcal{S}_{a})^{2}+(\partial _{k}\mathcal{P}_{a})^{2}
+(\partial _{k}\mathcal{R}_{ab})^{2}\right)   
 +2J_{s}^{-}\left( \sum (\partial _{k}\mathcal{S}_{a})^{2}+(\partial _{k}
\mathcal{P}_{z})^{2}+(\partial _{k}\mathcal{R}_{az})^{2}\right)   \notag \\
& +\rho _{\phi }\left[\epsilon _{\text{cap}}(\mathcal{P}_{z})^{2}-2\epsilon_{X}^{-}
\left( \sum (\mathcal{S}_{a})^{2}
+(\mathcal{R}_{az})^{2}\right)    -(\Delta _{\text{Z}}\mathcal{S}_{z}+\Delta _{\text{SAS}}\mathcal{P}_{x}
+\Delta _{\text{bias}}\mathcal{P}_{z})\right], 
\label{su4effectivehamiltonian1}
\end{align}
where $\rho_{\Phi }=\rho _{0}/\nu $ is the density of states, and
\begin{align}
J_s&=\frac{1}{16\sqrt{2\pi}}E^0_{\text{C}},\quad
J^d_s=J_s \left[
-\sqrt{\frac{2}{\pi}}\frac{d}{\ell _{B}}+
\left(
1+\frac{d^2}{\ell _{B}^2}\right)
e^{d^2/2\ell _{B}^2}\text{erfc}\left(d/\sqrt{2}\ell _{B}  \right)
\right],\notag\\
J_{s}^{\pm }&=\frac{1}{2}(J_{s}\pm J_{s}^{d}),\notag\\
\epsilon_X&=\frac{1}{2}\sqrt{\frac{\pi}{2}}E^0_{\text{C}}, \quad
\epsilon_X^\pm=\frac{1}{2}\left[ 
1\pm  e^{d^2/2\ell _{B}^2}\text{erfc}\left(d/\sqrt{2}\ell _{B} \right)
\right]\epsilon_X,\quad
\epsilon _{D}^{-}=\frac{d}{4\ell _{B}}E^0_\text{C}, \notag\\
\epsilon _{\text{cap}}&=4\epsilon _{D}^{-}-2\epsilon _{X}^{-}, 
\end{align}
with
\begin{equation}
E^0_{\text{C}}=\frac{e^2}{4\pi\epsilon \ell _{B}}.  
\end{equation}
This Hamiltonian is valid at $\nu =1, 2$ and $3$.

It should be noted that all potential terms vanish in the SU(4)-invariant
limit, where perturbative excitations are gapless. They are the
NG modes associated with spontaneous breaking of  SU(4)
symmetry. They get gapped in the actual system, since  SU(4) symmetry is
explicitly broken. Nevertheless, we call them the NG modes.

\section{Bilayer quantum Hall system at $\nu=1$}
In this section, we first show the ground state structure and the associated
 NG modes. We then show the interlayer phase coherence, the associated Josephson supercurrent, and its effect on the Hall resistance, 
in the limit $\Delta _{\text{SAS}}\rightarrow0$.




\subsection{Ground state structure}
\label{nu1groundstatestructure}  
We introduce the $\text{CP}^{3}$ field based on the composite boson theory.   
An electron is converted into a composite boson by acquiring a flux quantum in the QH state.
The CP$^{3}$ field emerges when composite bosons undergo
Bose-Einstein condensation. 
The dimensionless SU(4) isospin densities are given by\cite{Ezawa:2008ae}:
\protect\begin{align}
\mathcal{S}_{a}(\boldsymbol{x})& =\frac{1}{2}\boldsymbol{n}^{\dagger }%
\protect\tau _{a}^{\text{spin}}\boldsymbol{n},  \nonumber \\
\mathcal{P}_{a}(\boldsymbol{x})& =\frac{1}{2}\boldsymbol{n}^{\dagger }%
\protect\tau _{a}^{\text{ppin}}\boldsymbol{n},  \nonumber \\
\mathcal{R}_{ab}(\boldsymbol{x})& =\frac{1}{2}\boldsymbol{n}^{\dagger }%
\protect\tau _{a}^{\text{spin}}\protect\tau _{b}^{\text{ppin}}\boldsymbol{n},
\protect\end{align}%
where $\boldsymbol{n}$ is the CP$^{3}$ field of the form 
$\boldsymbol{n}(\boldsymbol{x})=\left( n^{\text{f}\uparrow },n^{\text{f}\downarrow },n^{\text{b}\uparrow
},n^{\text{b}\downarrow }\right) ^{t}$.

The ground state at the imbalanced configuration $\sigma_0$ is given by
\begin{equation}
(n^{\text{B}\uparrow}_g,n^{\text{B}\downarrow}_g,
n^{\text{A}\uparrow}_g,n^{\text{A}\downarrow}_g)
=(1,0,0,0), 
\end{equation}
in the bonding-antibonding representation, which reads
\begin{align}
\left( 
\begin{array}{c}
n^{\text{f}\uparrow}_g  \\
n^{\text{f}\downarrow}_g  \\
n^{\text{b}\uparrow}_g   \\
n^{\text{b}\downarrow}_g   \\
\end{array} 
\right)=\frac{1}{\sqrt{2}}
\left( 
\begin{array}{cccc}
\sqrt{1+\sigma_0} & 0 & \sqrt{1-\sigma_0} & 0 \\
0 & \sqrt{1+\sigma_0} & 0 & \sqrt{1-\sigma_0}  \\
\sqrt{1-\sigma_0} & 0 & -\sqrt{1+\sigma_0} & 0   \\
0 & \sqrt{1-\sigma_0} & 0 & -\sqrt{1+\sigma_0}   \\ 
\end{array} 
\right)
\left( 
\begin{array}{c}
n^{\text{B}\uparrow}_g  \\
n^{\text{B}\downarrow}_g  \\
n^{\text{A}\uparrow}_g   \\
n^{\text{A}\downarrow}_g   \\
\end{array} 
\right)=
\left( 
\begin{array}{c}
\sqrt{\frac{1+\sigma_0}{2}}  \\
0  \\
\sqrt{\frac{1-\sigma_0}{2}}   \\
0  \\
\end{array} 
\right),\label{groundstatenu1cp3} 
\end{align}
in the layer representation. The ground-state values of the isospin fields are
\begin{align}
\mathcal{S}^g_a=\frac{1}{2}\delta_{az}, \quad 
\mathcal{P}^g_a=\frac{1}{2}\left(\sqrt{1-\sigma_0^2}\delta_{ax}+\sigma_0\delta_{az}\right),\quad
\mathcal{R}^g_{ab}=\frac{1}{2}\delta_{az}
\left(\sqrt{1-\sigma_0^2}\delta_{bx}+\sigma_0\delta_{bz}\right),\label{nu1orderparametes}
\end{align}
 all others being zero, giving a unique phase. The residual symmetry keeping the ground state invariant is U(3). 
Thus, the symmetry-breaking pattern is SU(4)$\rightarrow$U(3). The target space is the coset space
\begin{equation}
\text{CP}^3=\text{SU}(4)/\text{U}(3)
=\text{U}(4)/[\text{U}(1)\otimes\text{U}(3)],\label{nu1ssbpattern}
\end{equation}
which is the complex projective (CP) space. 

\subsection{Effective Hamiltonian for the NG modes at $\nu=1$}
\label{effectivehamiltoniannu1} 
\begin{figure}[t]
\begin{center}    
\centering
\includegraphics[width=0.9\textwidth]{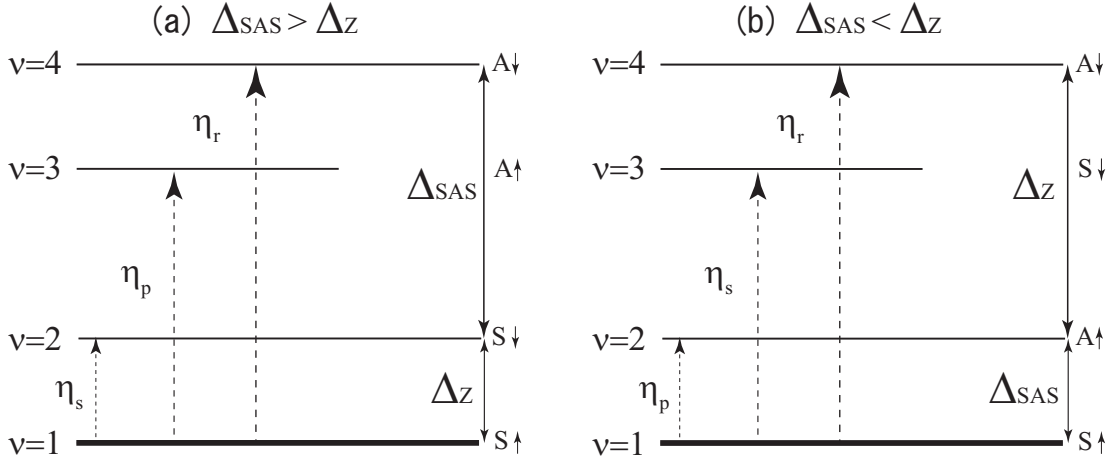}
\end{center}
\caption{  The lowest Landau level contains four energy levels corresponding to the two layers and the two spin states. 
They are shown in (a) for $\Delta_{\text{SAS}}>\Delta_{\text{Z}}$ and (b) for $\Delta_{\text{SAS}}<\Delta_{\text{Z}}.$
The lowest-energy level consists of up-spin symmetric states in the balanced configurations, and is filled at $\nu=1$. 
It is the spin-ferromagnet and pseudospin-ferromagnet state. Small fluctuations are NG modes 
$\eta_{\text{s}}$, $\eta_{\text{p}}$, and $\eta_{\text{r}}$.}
\label{nu1ngspectrum} 
\end{figure}  
From the previous subsection, we see that the symmetry-breaking pattern is given by
\eqref{nu1ssbpattern}, and therefore three complex NG modes emerge, which are described by the $\text{CP}^3$ fields.

We analyze the perturbative excitations around the ground state. 
We parameterize the bonding-antibonding state as
\begin{equation}
n^{\text{B}\uparrow}=\sqrt{1-|\eta_{\text{s}}|^2-|\eta_{\text{p}}|^2-|\eta_{\text{r}}|^2}, \quad
n^{\text{B}\downarrow}=\eta_{\text{s}}, \quad 
n^{\text{A}\uparrow}=\eta_{\text{p}}, \quad 
n^{\text{A}\downarrow}=\eta_{\text{r}}, 
\end{equation}
requiring the commutation relations
\begin{equation}
\left[\eta_i(\boldsymbol{x}),\eta_j^\dagger(\boldsymbol{y})\right]=
\rho_0^{-1}\delta_{ij}\delta(\boldsymbol{x}-\boldsymbol{y}),
\end{equation}
in order to satisfy the SU(4) algebraic relation.
$\eta_{\text{s}}$ describes the spin wave, $\eta_{\text{p}}$ the pseudospin wave, 
and $\eta_{\text{r}}$ the $R$-spin wave connecting the ground state to the highest level in the  
lowest level (Fig. \ref{nu1ngspectrum}).  
The layer field reads
\begin{align}
\left( 
\begin{array}{c}
n^{\text{f}\uparrow}  \\
n^{\text{f}\downarrow}  \\
n^{\text{b}\uparrow}   \\
n^{\text{b}\downarrow}   \\
\end{array} 
\right)=\frac{1}{\sqrt{2}}
\left( 
\begin{array}{cccc}
\sqrt{1+\sigma_0} & 0 & \sqrt{1-\sigma_0} & 0 \\
0 & \sqrt{1+\sigma_0} & 0 & \sqrt{1-\sigma_0}  \\
\sqrt{1-\sigma_0} & 0 & -\sqrt{1+\sigma_0} & 0   \\
0 & \sqrt{1-\sigma_0} & 0 & -\sqrt{1+\sigma_0}   \\ 
\end{array} 
\right)
\left( 
\begin{array}{c}
n^{\text{B}\uparrow}  \\
n^{\text{B}\downarrow}  \\
n^{\text{A}\uparrow}   \\
n^{\text{A}\downarrow}   \\
\end{array} 
\right).
\end{align}
Expanding 
\begin{equation}
(n^{\text{B}\uparrow}, n^{\text{B}\downarrow},  
n^{\text{A}\uparrow},n^{\text{A}\downarrow})=(1,\eta_{\text{s}},\eta_{\text{p}},\eta_{\text{r}})+\cdots, 
\end{equation}
for small fluctuations around the ground state, we obtain 
\begin{align}
n^{\text{f}\uparrow}&=\sqrt{\frac{1+\sigma_0}{{2}}}\left(1-\frac{1}{2}(|\eta_{\text{s}}|^2+|\eta_{\text{p}}|^2+|\eta_{\text{r}}|^2)\right)+\eta_{\text{p}}\sqrt{\frac{1-\sigma_0}{{2}}}, \quad
n^{\text{f}\downarrow}=\eta_{\text{s}}\sqrt{\frac{1+\sigma_0}{{2}}}+\eta_{\text{r}}\sqrt{\frac{1-\sigma_0}{{2}}},\notag\\
n^{\text{b}\uparrow}&=\sqrt{\frac{1-\sigma_0}{{2}}}\left(1-\frac{1}{2}(|\eta_{\text{s}}|^2+|\eta_{\text{p}}|^2
+|\eta_{\text{r}}|^2)\right)-\eta_{\text{p}}\sqrt{\frac{1+\sigma_0}{{2}}}, \quad
n^{\text{b}\downarrow}=\eta_{\text{s}}\sqrt{\frac{1-\sigma_0}{{2}}}-\eta_{\text{r}}\sqrt{\frac{1+\sigma_0}{{2}}}.
\label{linearcp31}
\end{align} 
We then set  
\begin{equation}
\eta_i(\boldsymbol{x})=\frac{\sigma_i(\boldsymbol{x})+i\vartheta_i(\boldsymbol{x})}{2},
\end{equation}
where $\rho_0\sigma_i(\boldsymbol{x})$ is the number density excited from the ground state to the $i$th level
designated by \eqref{linearcp31}, and $\vartheta_i(\boldsymbol{x})$ is the conjugate phase field, 
satisfying the commutation relation
\begin{equation}
\frac{\rho_0}{2}\left[\sigma_i(\boldsymbol{x}),\vartheta_j(\boldsymbol{y})\right]=
i\delta_{ij}\delta(\boldsymbol{x}-\boldsymbol{y}). 
\end{equation} 
We express the isospin field in terms of the $\text{CP}^3$ field \eqref{linearcp31},
\begin{align}
&2\mathcal{S}_a=\left(\sigma_{\text{s}}+\frac{1}{2}\left(\sigma_{\text{p}}\sigma_{\text{r}}+\vartheta_{\text{p}}\vartheta_{\text{r}}\right),\vartheta_{\text{s}}+\frac{1}{2}\left(\sigma_{\text{p}}\vartheta_{\text{r}}-\vartheta_{\text{p}}\sigma_{\text{r}}\right),1-2|\eta_{\text{s}}|^2-2|\eta_{\text{r}}|^2\right),\notag\\
&2\mathcal{P}_a=\left(p_x(\text{s},\text{p},\text{r}),-\vartheta_{\text{p}}-\frac{1}{2}(\sigma_{\text{s}}\vartheta_{\text{r}}-\vartheta_{\text{s}}\sigma_{\text{r}}),p_z(\text{s},\text{p},\text{r})\right),\notag\\
&2\mathcal{R}_{xa}=\left(r_{xx}(\text{s},\text{p},\text{r}),-\vartheta_{\text{r}}+\frac{1}{2}(\sigma_{\text{p}}\vartheta_{\text{s}}-\vartheta_{\text{p}}\sigma_{\text{s}}),
r_{xz}(\text{s},\text{p},\text{r})\right),\notag\\
&2\mathcal{R}_{ya}=\left(r_{yx}(\text{s},\text{p},\text{r}),\sigma_{\text{r}}-\frac{1}{2}(\sigma_{\text{s}}\sigma_{\text{p}}+\vartheta_{\text{s}}\vartheta_{\text{p}}),
r_{yz}(\text{s},\text{p},\text{r})\right),\notag\\
&2\mathcal{R}_{za}=\left({r_{zy}(\text{s},\text{p},\text{r})},-\vartheta_{\text{p}}+\frac{1}{2}(\sigma_{\text{s}}\vartheta_{\text{r}}-\vartheta_{\text{s}}\sigma_{\text{r}}),{r_{zz}(\text{s},\text{p},\text{r})}\right),
\end{align} 
with 
\begin{align}  
&p_x(\text{s},\text{p},\text{r})=\sqrt{1-\sigma^2_0}-\sigma_0\sigma_{\text{p}}-2\sqrt{1-\sigma^2_0}
\left(|\eta_{\text{p}}|^2+|\eta_{\text{r}}|^2\right)-\frac{\sigma_0}{2}(\sigma_{\text{s}}\sigma_{\text{r}}+\vartheta_{\text{s}}\vartheta_{\text{r}}),\notag\\  
&p_z(\text{s},\text{p},\text{r})={\sigma_0}+\sqrt{1-\sigma^2_0}\sigma_{\text{p}}
-2\sigma_0\left(|\eta_{\text{p}}|^2+|\eta_{\text{r}}|^2\right)
+\frac{\sqrt{1-\sigma^2_0}}{2}(\sigma_{\text{s}}\sigma_{\text{r}}+\vartheta_{\text{s}}\vartheta_{\text{r}}),\nonumber\\
&r_{xx}(\text{s},\text{p},\text{r})=\sqrt{1-\sigma^2_0}\sigma_{\text{s}}-\sigma_0\sigma_{\text{r}}-\frac{\sigma_0}{2}\left(\sigma_{\text{s}}\sigma_{\text{p}}+\vartheta_{\text{s}}\vartheta_{\text{p}}\right)-\frac{\sqrt{1-\sigma_0^2}}{2}\left(\sigma_{\text{p}}\sigma_{\text{r}}+\vartheta_{\text{p}}\vartheta_{\text{r}}\right),\nonumber\\
&r_{yx}(\text{s},\text{p},\text{r})=\sqrt{1-\sigma^2_0}\vartheta_{\text{s}}-\sigma_0\vartheta_{\text{r}}+\frac{\sigma_0}{2}\left(\sigma_{\text{s}}\vartheta_{\text{p}}-\vartheta_{\text{s}}\sigma_{\text{p}}\right)-\frac{\sqrt{1-\sigma_0^2}}{2}\left(\sigma_{\text{p}}\vartheta_{\text{r}}-\vartheta_{\text{p}}\sigma_{\text{r}}\right),\nonumber\\
&r_{xz}(\text{s},\text{p},\text{r})=\sigma_0\sigma_{\text{s}}+\sqrt{1-\sigma^2_0}\sigma_{\text{r}}-\frac{\sigma_0}{2}\left(\sigma_{\text{p}}\sigma_{\text{r}}+\vartheta_{\text{p}}\vartheta_{\text{r}}\right)+\frac{\sqrt{1-\sigma_0^2}}{2}\left(\sigma_{\text{s}}\sigma_{\text{p}}+\vartheta_{\text{s}}\vartheta_{\text{p}}\right),\nonumber\\
&r_{yz}(\text{s},\text{p},\text{r})=\sigma_0\vartheta_{\text{s}}+\sqrt{1-\sigma^2_0}\vartheta_{\text{r}}-\frac{\sigma_0}{2}\left(\sigma_{\text{p}}\vartheta_{\text{r}}-\vartheta_{\text{p}}\sigma_{\text{r}}\right)-\frac{\sqrt{1-\sigma_0^2}}{2}\left(\sigma_{\text{s}}\vartheta_{\text{p}}-\vartheta_{\text{s}}\sigma_{\text{p}}\right),\nonumber\\
&{r_{zx}(\text{s},\text{p},\text{r})=\sqrt{1-\sigma^2_0}-\sigma_0\sigma_{\text{p}}-2\sqrt{1-\sigma^2_0}
\left(|\eta_{\text{p}}|^2+|\eta_{\text{s}}|^2\right)+\frac{\sigma_0}{2}(\sigma_{\text{s}}\sigma_{\text{r}}+\vartheta_{\text{s}}\vartheta_{\text{r}})},\notag\\
&{r_{zz}(\text{s},\text{p},\text{r})={\sigma_0}+\sqrt{1-\sigma^2_0}\sigma_{\text{p}}
-2\sigma_0\left(|\eta_{\text{p}}|^2+|\eta_{\text{s}}|^2\right)
-\frac{\sqrt{1-\sigma^2_0}}{2}(\sigma_{\text{s}}\sigma_{\text{r}}+\vartheta_{\text{s}}\vartheta_{\text{r}})}.
\end{align}
Substituting these into \eqref{su4effectivehamiltonian1}, we obtain the effective Hamiltonian
\begin{equation}
\int d^2 k\mathcal{H}_{\text{eff}}=\int d^2 k\mathcal{H}_{\text{ppin}}+\int d^2 k\mathcal{H}_{\text{mix}},
\end{equation}
with
\begin{align}
\mathcal{H}_{\text{ppin}}&=\frac{(1-\sigma_0^2)J_s+\sigma_0^2J^d_s}{2}(\partial_k\sigma_{\text{p}})^2
+\frac{\rho_0}{4}\left[\epsilon^{\nu=1}_{\text{cap}}(1-\sigma_0^2)
+\frac{\Delta_{\text{SAS}}}{\sqrt{1-\sigma_0^2}} \right]\sigma_{\text{p}}^2\notag\\
&+\frac{1}{2}J^d_s(\partial_k\vartheta_{\text{p}})^2+\frac{\rho_0}{4}\frac{\Delta_{\text{SAS}}}{\sqrt{1-\sigma_0^2}}\vartheta_{\text{p}}^2,\label{nu1ppinhamiltonian1}\\
\mathcal{H}_{\text{mix}}&=\frac{J^+_s+\sigma_0J^-_s}{2}\left[(\partial_k\sigma_1)^2+(\partial_k\vartheta_1)^2\right]
+\frac{\rho_0}{4}\left(
\Delta_{\text{Z}}+\frac{1}{2}\Delta_{\text{SAS}}\frac{\sqrt{1-\sigma_0}}{\sqrt{1+\sigma_0}}
\right)\left[\sigma_1^2+\vartheta_1^2\right]\notag\\
&+\frac{J^+_s-\sigma_0J^-_s}{2}\left[(\partial_k\sigma_2)^2+(\partial_k\vartheta_2)^2\right]
+\frac{\rho_0}{4}\left(
\Delta_{\text{Z}}+\frac{1}{2}\Delta_{\text{SAS}}\frac{\sqrt{1-\sigma_0}}{\sqrt{1+\sigma_0}}
\right)\left[\sigma_2^2+\vartheta_2^2\right]\notag\\
&{-\frac{\rho_0}{4}\Delta_{\text{SAS}}(\sigma_1\sigma_2+\vartheta_1\vartheta_2)}
\label{mixnu1hamiltonian},
\end{align}
where we change the variables in \eqref{mixnu1hamiltonian} as
\begin{equation}
\eta_{\text{s}}=\sqrt{\frac{{1+\sigma_0}}{2}}\eta_1+\sqrt{\frac{{1-\sigma_0}}{2}}\eta_2,\quad
\eta_{\text{r}}=\sqrt{\frac{{1-\sigma_0}}{2}}\eta_1-\sqrt{\frac{{1+\sigma_0}}{2}}\eta_2,
\end{equation}
and  $\Delta_{\text{bias}}$ and $\epsilon^{\nu=1}_{\text{cap}}$ are given by
\begin{align}
\Delta_{\text{bias}}&=\frac{\sigma_0 }{\sqrt{1-\sigma_0^2}}\Delta_{\text{SAS}}+\sigma_0 \epsilon^{\nu=1}_{\text{cap}} ,\\
\epsilon^{\nu=1}_{\text{cap}}&=4(\epsilon^-_D-\epsilon^-_X ),
\end{align}
respectively. 
The pseudospin mode is decoupled from other modes, and from \eqref{nu1ppinhamiltonian1} 
we have coherence lengths of the interlayer phase field $\vartheta_{\text{p}}$ and the imbalanced field $\sigma_{\text{p}}$ 
\begin{align}
\xi^\vartheta_{\text{ppin}}&=2l_B\sqrt{\frac{\pi\sqrt{1-\sigma_0^2}J^d_s}{\Delta_{\text{SAS}}}},\notag\\
\xi^\sigma_{\text{ppin}}&=2l_B\sqrt{\frac{\pi\left[(1-\sigma_0^2)J_s+\sigma_0^2J^d_s
\right]}
{\epsilon^{\nu=1}_{\text{cap}}(1-\sigma_0^2)+\Delta_{\text{SAS}}/\sqrt{1-\sigma_0^2}}}.
\end{align}
The $\vartheta_{\text{p}}$ mode is gapless for $\Delta_{\text{SAS}}=0$, though the $\sigma_{\text{p}}$ mode is gapful due 
to the capacitance term $\epsilon^{\nu=1}_{\text{cap}}$.

On the other hand, from \eqref{mixnu1hamiltonian} for $\Delta_{\text{SAS}}=0$, 
the two modes  $\eta_1$ and $\eta_2$ are decoupled.
There exist no gapless modes in the Hamiltonian \eqref{mixnu1hamiltonian} provided $\Delta_{\text{Z}}\neq0$. 
\subsection{Effective Hamiltonian for the NG modes in the limit $\Delta_{\text{SAS}}\rightarrow 0$}
\label{effectivehamiltoniannu1Dsas0}
We concentrate solely on the gapless mode in the limit $\Delta_{\text{SAS}}\rightarrow 0$, 
since we are interested in the interlayer coherence in this system.
We now analyze the nonperturbative
phase-coherent phenomena, where the phase field $\vartheta(\boldsymbol{x})$
is essentially classical and may become very large.
We parameterize the $\text{CP}^3$ field as 
\begin{align}
\left( 
\begin{array}{c}
n^{\text{f}\uparrow}(\boldsymbol{x})  \\
n^{\text{f}\downarrow}(\boldsymbol{x})  \\
n^{\text{b}\uparrow}(\boldsymbol{x})   \\
n^{\text{b}\downarrow}(\boldsymbol{x})   \\
\end{array} 
\right)=\frac{1}{\sqrt{2}} 
\left( 
\begin{array}{c}
e^{i\vartheta(\boldsymbol{x})/2}\sqrt{1+\sigma(\boldsymbol{x})} \\
0 \\
e^{-i\vartheta(\boldsymbol{x})/2}\sqrt{1-\sigma(\boldsymbol{x})} \\
0 \\
\end{array} 
\right).
\label{nu1cp3}
\end{align}
Then  the isospin fields are expressed as 
\begin{align} 
\mathcal{S}_{z}(\boldsymbol{x})&=\frac{1}{2},\quad
\mathcal{P}_{z}(\boldsymbol{x})=\mathcal{R}_{zz}(\boldsymbol{x})
=\frac{1}{2}\sigma(\boldsymbol{x}),\nonumber\\
\mathcal{P}_{x}(\boldsymbol{x})&=\mathcal{R}_{zx}(\boldsymbol{x})=
\frac{1}{2}\sqrt{1-\sigma^2(\boldsymbol{x})}\cos\vartheta(\boldsymbol{x}),\quad
\mathcal{P}_{y}(\boldsymbol{x})=\mathcal{R}_{zy}(\boldsymbol{x})=
-\frac{1}{2}\sqrt{1-\sigma^2(\boldsymbol{x})}\sin\vartheta(\boldsymbol{x}),
\label{isospinnu1} 
\end{align} 
with all others being zero. 
From \eqref{isospinnu1} we obtain the effective Hamiltonian 
\begin{align} 
\mathcal{H}_{\text{eff}}&=\frac{J^d_s}{2}(1-\sigma^2(\boldsymbol{x}))(\partial_k\vartheta(\boldsymbol{x}))^2
+\frac{1}{2}\left( J_s+\frac{\sigma^2(\boldsymbol{x})}{1-\sigma^2(\boldsymbol{x})}J_s^d\right)(\partial_k\sigma(\boldsymbol{x}))^2\notag\\
&+\frac{\rho_0\epsilon^{\nu=1}_{\text{cap}}}{4}(\sigma(\boldsymbol{x})-\sigma_0)^2 
-\frac{\rho_0\Delta_{\text{SAS}}}{2}
\left(
\sqrt{1-\sigma^2(\boldsymbol{x})}\cos\vartheta(\boldsymbol{x})+\frac{\sigma_0}{\sqrt{1-\sigma^2_0}}\sigma(\boldsymbol{x})  
\right).\label{nu1hamiltonian}
\end{align}
The canonical commutation relation is given by  
\begin{equation}
\frac{\rho_0}{2}\left[
\sigma(\boldsymbol{x}),\vartheta(\boldsymbol{x})
\right]=i\delta(\boldsymbol{x}-\boldsymbol{y}).\label{nu1ccr}
\end{equation}
From \eqref{nu1hamiltonian} and \eqref{nu1ccr}, the Heisenberg equations of motion can be
calculated as
\begin{align}
\hbar\partial_t\vartheta&=\frac{2}{\rho_0}\partial_k(J^\sigma_{\text{s}}\partial_k\sigma)+\frac{2J^d_s}{\rho_0}\sigma 
\left[
(\partial_k\vartheta)^2-\frac{1}{1-\sigma^2}(\partial_k\sigma)^2
\right]\nonumber\\ 
&-\epsilon^{\nu=1}_{\text{cap}}(\sigma-\sigma_0)
-\frac{\sigma\cos\vartheta}{\sqrt{1-\sigma^2}}\Delta_{\text{SAS}}+\frac{\sigma_0}{\sqrt{1-\sigma^2_0}}\Delta_{\text{SAS}},\label{nu1heom1}\\
\hbar\partial_t\sigma&=-\frac{2}{\rho_0}\partial_k(J^\vartheta_{\text{s}}\partial_k\vartheta)+\Delta_{\text{SAS}}\sqrt{1-\sigma^2}\sin\vartheta,
\label{nu1heom2}
\end{align}
with 
\begin{equation}
J^\vartheta_{\text{s}}=(1-\sigma^2)J^d_s, \quad 
J^\sigma_{\text{s}}=J_s+\frac{\sigma^2}{1-\sigma^2}J_s^d.
\end{equation}
\subsection{Josephson supercurrents }
\label{josephsonsupercurrentsnu1}
We now study the electric Josephson supercurrent carried by the gapless mode 
$\vartheta (\boldsymbol{x})$. In general, the total current  consists
of three types of current, the Josephson in-plane current
$\mathcal{J}^{\text{Jos}}_i$, the Josephson tunneling current 
$\mathcal{J}^{\text{Jos}}_z$, which is proportional to  $\Delta_{\text{SAS}}$, 
 and the Hall current $\mathcal{J}^{\text{Hall}}_i$. 
What has been
argued in \cite{Ezawa:2012epjb} is that in the case of $\nu=1$, 
there exists an interlayer voltage $V_{\text{junc}}$ and thus no dissipationless
 $\mathcal{J}^{\text{Jos}}_z$
exists, when $\sigma_0\neq0$. 
 On the other hand, the Josephson
in-plane current, which is dissipationless does exist, even
for $\sigma_0\neq0$. Here, we assume the sample parameter
$\sigma_0\neq0$ and  $\Delta_{\text{SAS}}=0$ 
so that there is no dissipationless tunneling current $\mathcal{J}^{\text{Jos}}_z$ between
the two layers.

The electron densities are 
$\rho_{e}^{\text{f}(\text{b})}=-{e\rho_{0}}\left( 1/2\pm \mathcal{P}_{z}\right) 
=-{e\rho_{0}}\left( 1\pm \sigma (\boldsymbol{x})\right) /2$ on each layer. 
Taking the time derivative and using \eqref{nu1heom2} we find 
\begin{equation}
\partial_{t}\rho_{e}^{\text{f}}=-\partial_{t}\rho_{e}^{\text{b}}
=\frac{eJ_s^{\vartheta}}{\hbar}\nabla^{2}\vartheta (\boldsymbol{x}).
\label{continuityequation1}
\end{equation}
The time derivative of the charge is associated with the current via the
continuity equation, $\partial_{t}\rho_{e}^{\text{f}(\text{b})}
=\partial_{i}\mathcal{J}_{i}^{\text{f}(\text{b})}$. We thus identify 
$\mathcal{J}_{i}^{\text{f(b)}}=\pm \mathcal{J}_{i}^{\text{Jos}}(\boldsymbol{x})+$constant, 
where
\begin{equation}
\mathcal{J}_{i}^{\text{Jos}}(\boldsymbol{x})\equiv \frac{eJ_s^{\vartheta}}{\hbar}
\partial_{i}\vartheta (\boldsymbol{x}).  \label{phasecurrent}
\end{equation} 
Consequently, the current $\mathcal{J}_{i}^{\text{Jos}}(\boldsymbol{x})$
flows when there exists inhomogeneity in the phase $\vartheta (\boldsymbol{x})$. 
Such a current is precisely the Josephson supercurrent. Indeed, it is a
supercurrent because the coherent mode exhibits a linear dispersion
relation. 
\subsection{Quantum Hall effects}
\label{qhenu1}
Let us inject the current $\mathcal{J}_{\text{in}}$ into the $x$ direction
of the bilayer sample, and assume the system to be homogeneous in the $y$
direction (Fig.\ref{nu1spincurrentfigure}). This creates the electric field 
$E_{y}^{\text{f(b)}}$ so that the Hall current flows into the $x$-direction.
A bilayer system consists of the two layers and the volume between them. The
Coulomb energy in the volume is minimized\cite{Ezawa:2007nj} by the
condition $E_{y}^{\text{f}}=E_{y}^{\text{b}}$. We thus impose
$E_{y}^{\text{f}}=E_{y}^{\text{b}}\equiv E_{y}$. 
The current is the sum of the Hall current and the Josephson current, 
\begin{equation}
\mathcal{J}_{x}^{\text{f}}(x)=\frac{\nu}{R_{\text{K}}}
\frac{\rho_{0}^{\text{f}}}{\rho_{0}}E_{y}+\mathcal{J}_{x}^{\text{Jos}},
\quad \mathcal{J}_{x}^{\text{b}}(x)=\frac{\nu}{R_{\text{K}}}\frac{\rho_{0}^{\text{b}}}
{\rho_{0}}E_{y}-\mathcal{J}_{x}^{\text{Jos}},  \label{totalcurrent}
\end{equation}
with $R_{\text{K}}=2\pi \hbar /e^{2}$ the von Klitzing constant. We obtain
the standard Hall resistance when $\mathcal{J}_{x}^{\text{Jos}}=0$. That is, 
the emergence of the Josephson supercurrent is detected if the Hall
resistance becomes anomalous.

We apply these formulas to analyze the counterflow and drag experiments
since they occur without tunneling. In\ the counterflow experiment, the
current $\mathcal{J}_{\text{in}}$ is injected to the front layer and
extracted from the back layer at the same edge. Since there is no tunneling
we have 
$\mathcal{J}_{x}^{\text{b}}=-\mathcal{J}_{x}^{\text{f}}=-\mathcal{J}_{\text{in}}$. 
Hence, it follows from (\ref{totalcurrent}) that $E_{y}=0$, or
\begin{equation}
R_{xy}^{\text{f}}\equiv \frac{E_{y}^{\text{f}}}{\mathcal{J}_{x}^{\text{f}}}=0,
\qquad 
R_{xy}^{\text{b}}\equiv \frac{E_{y}^{\text{b}}}{\mathcal{J}_{x}^{\text{b}}}=0.  \label{counterflowanomalous}
\end{equation}
All the input current is carried by the Josephson supercurrent, 
$\mathcal{J}_{x}^{\text{Jos}}=\mathcal{J}_{\text{in}}$. It generates such an
inhomogeneous phase field that 
$\vartheta (\boldsymbol{x})=(\hbar/eJ_s^{\vartheta})\mathcal{J}_{\text{in}}x$.

On the other hand, in the drag experiment, since  interlayer-coherent
tunneling is absent, no current flows on the back layer, or $\mathcal{J}_{x}^{\text{b}}=0$. 
Hence, it follows from (\ref{totalcurrent}) that 
$\mathcal{J}_{\text{in}}=\mathcal{J}_{x}^{\text{f}}=(\nu /R_{\text{K}})E_{y}$, or 
\begin{equation}
R_{xy}^{\text{f}}\equiv {\frac{E_{y}^{\text{f}}}{\mathcal{J}_{x}^{\text{f}}}=
\frac{R_{\text{K}}}{\nu}},
 \label{draganomalous}
\end{equation}
A part of the input current is carried by the Josephson supercurrent, 
$\mathcal{J}_{x}^{\text{Jos}}=\frac{1}{2}(1-\sigma_{0})\mathcal{J}_{\text{in}}$.

\subsection{Spin Josephson supercurrents} 
\label{spinjosephsonsupercurrentsnu1}
\begin{figure}[t]
\begin{center}
\centering
\includegraphics[width=0.85\textwidth]{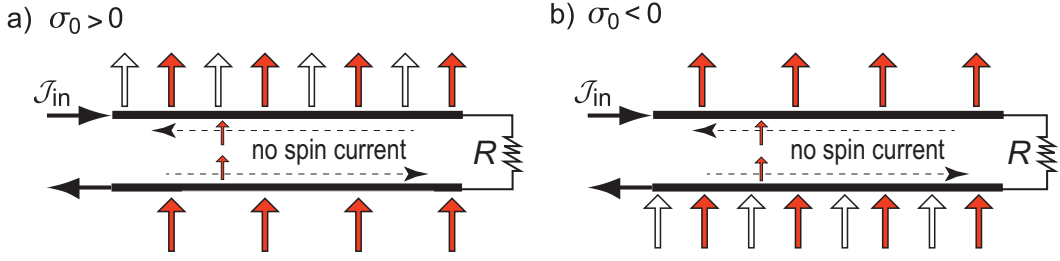}
\end{center}
\caption{  Schematic illustration of the spin supercurrent
flowing along the $x$ axis in the counterflow geometry for $\nu=1$  bilayer QH  system. 
 (a) In the $\nu=1$  bilayer QH  system for $\protect\sigma_{0}>0$, 
all spins are polarized into the positive $z$ axis. 
The interlayer phase
difference $\protect\vartheta (\boldsymbol{x})$ is created by feeding a
charge current $\mathcal{J}_{\text{in}}$ to the front layer, which also
drives the spin current. Electrons flow in each layer as indicated by the
dotted horizontal arrows.  
The direction of the spin current flowing in the front layer becomes
opposite to the direction of that flowing in the back layer,
and therefore no spin current flows as a whole. 
(b) In the $\nu=1$  QH bilayer system for $\protect\sigma_{0}<0$,
similar phenomena occur and therefore no spin current flows as a whole.}
\label{nu1spincurrentfigure}
\end{figure}
The spin density in each layer is defined by 
$\rho_{\alpha}^{\text{spin}}(\boldsymbol{x})
\equiv s_{\alpha}\psi_{\alpha}^{\dagger}\psi_{\alpha}$, 
where $s_{\alpha}=\frac{1}{2}\hbar $ 
for $\alpha =\text{f}\uparrow,\text{b}\uparrow$ and 
$s_{\alpha}=-\frac{1}{2}\hbar$ for $\alpha =\text{f}\downarrow,\text{b}\downarrow $.
By using the formula
\begin{align} 
\left( 
\begin{array}{c}
\rho_{\text{f}\uparrow} (\boldsymbol{x})  \\
\rho_{\text{f}\downarrow} (\boldsymbol{x})   \\
\rho_{\text{b}\uparrow} (\boldsymbol{x})   \\
\rho_{\text{b}\downarrow} (\boldsymbol{x})   \\
\end{array} 
\right)=
\frac{1}{4}\left( 
\begin{array}{cccc}
1 & 1 & 1 & 1 \\
1 & -1 & 1 & -1 \\
1 & 1 & -1 & -1 \\
1 & -1 & -1 & 1 \\
\end{array} 
\right)
\left( 
\begin{array}{c}
\rho_0  \\
2{S}_{z}(\boldsymbol{x})   \\
2{P}_{z}(\boldsymbol{x})   \\
2{R}_{zz}(\boldsymbol{x})   \\
\end{array} 
\right),
\end{align}
and \eqref{isospinnu1} we have
 \begin{align}
\left( 
\begin{array}{c}
\rho_{\text{f}\uparrow} (\boldsymbol{x})  \\
\rho_{\text{f}\downarrow} (\boldsymbol{x})   \\
\rho_{\text{b}\uparrow} (\boldsymbol{x})   \\
\rho_{\text{b}\downarrow} (\boldsymbol{x})   \\
\end{array} 
\right)=
\frac{\rho_0}{2}\
\left( 
\begin{array}{c}
1+\sigma(\boldsymbol{x})   \\
0  \\
1-\sigma(\boldsymbol{x})    \\
0   \\
\end{array} 
\right).
\end{align}
Then taking the time derivative for $\rho_\alpha$, we have
\begin{align}
\left( 
\begin{array}{c}
\partial_t\rho^{\text{spin}}_{\text{f}\uparrow} (\boldsymbol{x})  \\
\partial_t\rho^{\text{spin}}_{\text{f}\downarrow} (\boldsymbol{x})   \\
\partial_t\rho^{\text{spin}}_{\text{b}\uparrow} (\boldsymbol{x})   \\
\partial_t\rho^{\text{spin}}_{\text{b}\downarrow} (\boldsymbol{x})   \\
\end{array} 
\right)=
\frac{\hbar\rho_0}{4}\
\left( 
\begin{array}{c}
\partial_t\sigma(\boldsymbol{x})   \\
0  \\
-\partial_t\sigma(\boldsymbol{x})    \\
0   \\
\end{array} 
\right).
\end{align}
The time derivative of the spin is associated with the spin current
via the continuity equation (in this article we neglect the tunneling current): 
\begin{equation} 
\partial_t
\rho^{\text{Spin}}_{\alpha}(\boldsymbol{x})=
\partial_x \mathcal{J}_\alpha^{\text{Spin}}(\boldsymbol{x}),
\end{equation}
for each $\alpha.$ We thus identify 
\begin{align} 
\mathcal{J}_{\text{f}\uparrow}^{\text{Spin}}(\boldsymbol{x})&=-\mathcal{J}_{\text{b}\uparrow}^{\text{Spin}}(\boldsymbol{x})=
-\frac{J^{\vartheta}_s}{2}\partial_x\vartheta(\boldsymbol{x}), \quad \text{all others}=0.
\label{nu1spincurrent}
\end{align} 
Therefore from \eqref{nu1spincurrent} we see that 
the total spin current $\mathcal{J}^{\text{Spin}}\equiv\sum_\alpha\mathcal{J}_\alpha^{\text{Spin}}$
is zero, and therefore the spin Josephson supercurrent does not flow  at $\nu=1$ (Fig. \ref{nu1spincurrentfigure}).

\section{Bilayer quantum Hall system at $\nu=2$}

The standard Hall resistance is given by 
$R_{xy}^{\text{f}}=\frac{2}{\nu}R_{\text{K}}=R_{\text{K}}$ at $\nu =2$. 
On the other hand, it has been found experimentally \cite{Kellog1,Tutuc,Kellog2} that
$R_{xy}^{\text{f}}=R_{\text{K}}$ at $\nu =2$. 
It seems that the interlayer phase coherence together with the supercurrent 
does not develop at $\nu =2$. 
Note that the experiments \cite{Kellog1,Tutuc,Kellog2} were performed at the balance point $\sigma_0=0$. 
As we now demonstrate, the interlayer phase coherence develops only 
at the imbalance point $\sigma_0\neq0$ in the CAF phase.  

In this section, we first show the ground state structure and 
the NG modes for each phase. We then discuss 
the entangled spin-pseudospin phase coherence, 
the associated Josephson supercurrent and its effect on the Hall resistance 
in the CAF phase in the limit $\Delta _{\text{SAS}}\rightarrow0$. 

\subsection{Ground state structure}
\label{nu2groundstatestructure} 
It has been shown\cite{Ezawa:2005xi} at $\nu =2$
 that the order parameters, which are the classical isospin densities for the ground
state, are given in terms of two parameters $\alpha $ and $\beta $ as 
\begin{align} 
\mathcal{S}_{z}^{0}& =\frac{\Delta _{\text{Z}}}{\Delta _{0}}(1-\alpha ^{2})
\sqrt{1-\beta ^{2}},  \quad
\mathcal{P}_{x}^{0} =\frac{\Delta _{\text{SAS}}}{\Delta _{0}}\alpha ^{2}
\sqrt{1-\beta ^{2}},\ \ \mathcal{P}_{z}^{0}=\frac{\Delta _{\text{SAS}}}
{\Delta _{0}}\alpha ^{2}\beta ,  \notag \\
\mathcal{R}_{xx}^{0}& =-\frac{\Delta _{\text{SAS}}}
{\Delta _{0}}\alpha \sqrt{1-\alpha ^{2}}\beta,  \quad
\mathcal{R}_{yy}^{0} =-\frac{\Delta _{\text{Z}}}{\Delta _{0}}
\alpha \sqrt{1-\alpha ^{2}}\sqrt{1-\beta ^{2}},  \quad
\mathcal{R}_{xz}^{0} =\frac{\Delta _{\text{SAS}}}
{\Delta _{0}}\alpha \sqrt{1-\alpha ^{2}}\sqrt{1-\beta ^{2}},
\label{orderparameter1}
\end{align}
with all others being zero.
The parameters $\alpha $ and $\beta $, satisfying $|\alpha |\leq 1$
and $|\beta |\leq 1$, are determined by the variational equations as 
\begin{align} 
\Delta_{\text{Z}}^{2} &=\frac{\Delta _{\text{SAS}}^{2}}{1-\beta ^{2}}
-\frac{4\epsilon _{X}^{-}\left( \Delta _{0}^{2}-\beta ^{2}
\Delta _{\text{SAS}}^{2}\right) }{\Delta _{0}\sqrt{1-\beta ^{2}}}, \label{vareq1}  \\ 
\frac{\Delta _{\text{bias}}}{\beta \Delta _{\text{SAS}}} &=
\frac{4\left(\epsilon _{X}^{-}+2\alpha ^{2}(\epsilon _{\text{D}}^{-}
-\epsilon_{X}^{-})\right) }{\Delta _{0}}+\frac{1}{\sqrt{1-\beta ^{2}}},
\label{vareq2}
\end{align} 
where 
\begin{equation}
\Delta _{0}=\sqrt{\Delta _{\text{SAS}}^{2}\alpha ^{2}
+\Delta_{\text{Z}}^{2}(1-\alpha ^{2})(1-\beta ^{2})}.  \label{EqD0}
\end{equation}
As a physical variable it is more convenient to use the imbalance parameter
defined by
\begin{equation}
\sigma _{0}\equiv \mathcal{P}_{z}^{0}=\frac{\Delta _{\text{SAS}}}
{\Delta _{0}}\alpha ^{2}\beta,   \label{ImbalParam}
\end{equation}
instead of the bias voltage $\Delta _{\text{bias}}$. 
This is possible in the pseudospin and CAF phases. 
The bilayer system is
balanced at $\sigma _{0}=0$, while all electrons are in the front layer at 
$\sigma _{0}=1$, and in the back layer at $\sigma _{0}=-1$. 

There are three phases in the bilayer QH system at $\nu =2$. We discuss them
in terms of $\alpha $ and $\beta $.

First, when $\alpha =0$, it follows that $\mathcal{S}_{z}^{0}=1$, 
$\mathcal{P}_{a}^{0}=\mathcal{R}_{ab}^{0}=0$, since $\Delta _{0}=\Delta_{\text{Z}}\sqrt{1-\beta ^{2}}$.
Note that $\beta $ disappears from all formulas in \eqref{orderparameter1}.
This is the spin phase, which is characterized by the fact that the isospin
is fully polarized into the spin direction with  
\begin{equation}
\mathcal{S}_{z}^{0}=1,
\end{equation}
all others being zero. The spins in both layers point to the positive $z$
axis due to the Zeeman effect.

Second, when $\alpha =1$, it follows that $\mathcal{S}_{z}^{0}=0$ 
and $(\mathcal{P}_{x}^{0})^{2}+(\mathcal{P}_{z}^{0})^{2}=1$. This is the pseudospin
phase, which is characterized by the fact that the isospin is fully
polarized into the pseudospin direction with 
\begin{equation}
\mathcal{P}_{x}^{0}=\sqrt{1-\beta ^{2}},\qquad \mathcal{P}_{z}^{0}=\beta
=\sigma _{0},  \label{ppinorderparameters1}
\end{equation}
all the others being zero. 

For intermediate values of $\alpha $ ($0<\alpha <1$), not only the spin and
pseudospin but also some components of the residual spin are nonvanishing,
and we may control the density imbalance by applying a bias voltage as in
the pseudospin phase. It follows from \eqref{orderparameter1} that, as the system
goes away from the spin phase $(\alpha =0)$, the spins begin to cant
coherently and make antiferromagnetic correlations between the two layers.
Hence it is called the canted antiferromagnetic phase.

The interlayer phase coherence is an intriguing phenomenon in the bilayer QH system\cite{Ezawa:2008ae}.
Since it is enhanced in the limit $\Delta _{\text{SAS}}\rightarrow 0$,
it is interesting to also investigate the effective Hamiltonian in this limit
at $\nu=2$.  
We need to know how the parameters 
$\alpha $ and $\beta $ are expressed in terms of the physical variables.  
The solutions for \eqref{EqD0} are 
\begin{equation}
\beta =\pm \sqrt{1-\left( \frac{\Delta _{\text{SAS}}}{\Delta _{\text{Z}}}
\right) ^{2}}+O(\Delta _{\text{SAS}}^{4}),  \label{limitsolution1} 
\end{equation}
with 
\begin{equation}
\Delta _{0}\rightarrow \Delta _{\text{SAS}}+O(\Delta _{\text{SAS}}^{3}),
\label{limitsolution2}
\end{equation}
as we shall derive in \eqref{weakbeta2}.
 By using (\ref{ImbalParam}) we have 
\begin{equation}
\mathcal{P}_{z}^{0}=\sigma _{0}=\pm \alpha ^{2} 
+O(\Delta _{\text{SAS}}^{2}).
\label{ImbalCanted}
\end{equation}
The parameters $\alpha $ and $\beta $ are simple functions of the physical
variables $\Delta _{\text{SAS}}/\Delta _{\text{Z}}$ and $\sigma _{0}$ in the
limit $\Delta _{\text{SAS}}\rightarrow 0$. 

In particular, one of the layers becomes empty in the pseudospin phase 
and also near the pseudospin phase boundary in the CAF phase, 
since we have $\sigma _{0}\rightarrow \pm 1$ as $\alpha \rightarrow 1$. 
On the other hand, the bilayer system becomes balanced in the spin phase
and also near the spin phase boundary in the CAF phase,  
since we have $\sigma _{0}\rightarrow 0$ as $\alpha \rightarrow 0$.  

\subsection{Grassmannian approach} 
\label{grassmanniannu2}
We employ the Grassmannian formalism\cite{Hasebe} to make the
physical picture of this NG mode clearer and to construct a theory
which is valid nonperturbatively. The Grassmannian field $Z(\boldsymbol{x})$
consists of two $\text{CP}^{3}$ fields $\boldsymbol{n}_{1}(\boldsymbol{x})$
and $\boldsymbol{n}_{2}(\boldsymbol{x})$ at $\nu =2$, since there are two
electrons per one Landau site. Due to the Pauli exclusion principle they
should be orthogonal one to another. Hence, we require 
\begin{equation}
\boldsymbol{n}_{i}^{\dagger}(\boldsymbol{x})\cdot \boldsymbol{n}_{j}(\boldsymbol{x})
=\delta_{ij}, 
\end{equation}
with $i=1,2$. Using a set of two $\text{CP}^{3}$ fields
subject to this normalization condition we introduce a $4\times 2$ matrix
field, the Grassmannian field given by 
\begin{equation}
Z(\boldsymbol{x})=(\boldsymbol{n}_{1},\boldsymbol{n}_{2}), 
\end{equation}
obeying 
\begin{equation}
Z^{\dagger}Z=\boldsymbol{1}.  
\end{equation}
Though we have introduced two fields $\boldsymbol{n}_{1}(\boldsymbol{x})$ and 
$\boldsymbol{n}_{2}(\boldsymbol{x})$, we cannot distinguish them quantum mechanically since
they describe two electrons in the same Landau site. Namely, two fields 
$Z(\boldsymbol{x})$  and $Z^\prime(\boldsymbol{x})$ are indistinguishable
physically when they are related by a local U(2) transformation $U(\boldsymbol{x})$,
\begin{equation}
Z^\prime(\boldsymbol{x})=Z(\boldsymbol{x})U(\boldsymbol{x}).
\end{equation}
By identifying these two fields $Z(\boldsymbol{x})$  and $Z^\prime(\boldsymbol{x})$, 
the $4\times2$ matrix field $Z(\boldsymbol{x})$ takes values on the Grassmann manifold $\text{G}_{4,2}$
defined by 
\begin{equation}
\text{G}_{4,2}=\frac{\text{SU}(4)}{\text{U}(1)\otimes\text{SU}(2)\otimes\text{SU}(2)}.
\end{equation}
The field $Z(\boldsymbol{x})$ is no longer a set of two independent $\text{CP}^3$ fields. 
It is a new object, called the Grassmannian field, carrying eight real degrees of freedom.

The dimensionless SU(4) isospin densities are given by 
\begin{align}
&\mathcal{S}_{a}(\boldsymbol{x}) =\frac{1}{2}\text{Tr}\left[ Z^{\dagger
}\tau_{a}^{\text{spin}}Z\right] 
=\frac{1}{2}\sum_{i=1}^{2}\boldsymbol{n}_{i}^{\dagger}\tau_{a}^{\text{spin}}\boldsymbol{n}_{i},  \notag \\
&\mathcal{P}_{a}(\boldsymbol{x}) =\frac{1}{2}\text{Tr}\left[ Z^{\dagger
}\tau_{a}^{\text{ppin}}Z\right] 
=\frac{1}{2}\sum_{i=1}^{2}
\boldsymbol{n}_{i}^{\dagger}\tau_{a}^{\text{ppin}}\boldsymbol{n}_{i}, \notag\\
&\mathcal{R}_{ab}(\boldsymbol{x}) =\frac{1}{2}\text{Tr}\left[ Z^{\dagger
}\tau_{a}^{\text{spin}}\tau_{b}^{\text{ppin}}Z\right] =\frac{1}{2}
\sum_{i=1}^{2}\boldsymbol{n}_{i}^{\dagger}\tau_{a}^{\text{spin}}
\tau_{b}^{\text{ppin}}\boldsymbol{n}_{i},  \label{su4isospin2}
\end{align} 
where $\boldsymbol{n}_{i}$ consists of the basis 
$\boldsymbol{n}_{i}(\boldsymbol{x})=\left( n^{\text{f}\uparrow},
n^{\text{f}\downarrow},n^{\text{b}\uparrow},n^{\text{b}\downarrow}\right)^{t}$. 
The ground state is given by Eq. \eqref{orderparameter1}, 
which we express in terms of the two $\text{CP}^3$ fields $\boldsymbol{n}_{i}^g$.
It is straightforward to show that it is given by 
$\boldsymbol{n}_{i}^g=U\bar{\boldsymbol{n}}_{i}^g$ with
\begin{align}
U&=\text{exp}\left[
-\frac{i}{2}\tau^{\text{ppin}}_y(\theta_\beta+\frac{\pi}{2})
\right]
\text{exp}\left[
-\frac{i}{2}\tau^{\text{spin}}_x\tau^{\text{ppin}}_y\theta_\alpha
\right]
\text{exp}\left[
\frac{i}{2}\tau^{\text{spin}}_y\tau^{\text{ppin}}_x\theta_\delta
\right]\notag\\
&=
\left(   
\begin{array}{cccc}
\cos\frac{(2\theta_\beta+\pi)}{4}\cos\frac{\theta_\delta-\theta_\alpha}{2} & -\sin\frac{(2\theta_\beta+\pi)}{4}\sin\frac{\theta_\delta+\theta_\alpha}{2} & -\sin\frac{(2\theta_\beta+\pi)}{4}\cos\frac{\theta_\delta+\theta_\alpha}{2} & \cos\frac{(2\theta_\beta+\pi)}{4}\sin\frac{\theta_\delta-\theta_\alpha}{2}  \\
\sin\frac{(2\theta_\beta+\pi)}{4}\sin\frac{\theta_\delta-\theta_\alpha}{2} & \cos\frac{(2\theta_\beta+\pi)}{4}\cos\frac{\theta_\delta+\theta_\alpha}{2} & -\cos\frac{(2\theta_\beta+\pi)}{4}\sin\frac{\theta_\delta+\theta_\alpha}{2} & -\sin\frac{(2\theta_\beta+\pi)}{4}\cos\frac{\theta_\delta-\theta_\alpha}{2}  \\
\sin\frac{(2\theta_\beta+\pi)}{4}\cos\frac{\theta_\delta-\theta_\alpha}{2} & \cos\frac{(2\theta_\beta+\pi)}{4}\sin\frac{\theta_\delta+\theta_\alpha}{2} & \cos\frac{(2\theta_\beta+\pi)}{4}\cos\frac{\theta_\delta+\theta_\alpha}{2} & \sin\frac{(2\theta_\beta+\pi)}{4}\sin\frac{\theta_\delta-\theta_\alpha}{2}  \\
-\cos\frac{(2\theta_\beta+\pi)}{4}\sin\frac{\theta_\delta-\theta_\alpha}{2} & \sin\frac{(2\theta_\beta+\pi)}{4}\cos\frac{\theta_\delta+\theta_\alpha}{2} & -\sin\frac{(2\theta_\beta+\pi)}{4}\sin\frac{\theta_\delta+\theta_\alpha}{2} & \cos\frac{(2\theta_\beta+\pi)}{4}\cos\frac{\theta_\delta-\theta_\alpha}{2}  \\
\end{array} 
\right),
\label{su4rotation}
\end{align}
where $\theta_\alpha$, $\theta_\beta$, and $\theta_\delta$ are given by 
\begin{align}
 \cos \theta_\alpha&\equiv\sqrt{1-\alpha^2}, \quad
 \sin {\theta_\alpha}\equiv\alpha, \quad
 \cos {\theta_\beta}\equiv\sqrt{1-\beta^2}, \quad
 \sin {\theta_\beta}\equiv-\beta, \notag\\
 \cos \theta_\delta&\equiv\frac{\Delta_{Z}\sqrt{1-\beta^2}}{\Delta_0}\sqrt{1-\alpha^2}, \quad
 \sin \theta_\delta\equiv\frac{\Delta_{\text{SAS}}}{\Delta_0}\alpha,
\label{sincosdefinition} 
\end{align}
and 
\begin{equation}
\bar{\boldsymbol{n}}_{1}^g=(1,0,0,0)^t, \quad
\bar{\boldsymbol{n}}_{2}^g=(0,0,1,0)^t.
\end{equation}
We may introduce perturbative excitation modes   
$\eta_i$ by introducing the two $\text{CP}^3$ fields 
$\boldsymbol{n}_{i}=U\bar{\boldsymbol{n}}_{i}$ with 
\begin{equation}
\bar{\boldsymbol{n}}_{1}
=\left(\begin{array}{c}
1-\frac{1}{2}|\eta_1|^2-\frac{1}{2}|\eta_3|^2 \\ 
\eta_1 \\ 
-\frac{1}{2}\eta^\dagger_4\eta_1-\frac{1}{2}\eta^\dagger_2\eta_3 \\ 
\eta_3
\end{array}\right) ,
\quad \bar{\boldsymbol{n}}_{2}
=\left(\begin{array}{c}
-\frac{1}{2}\eta^\dagger_1\eta_4-\frac{1}{2}\eta^\dagger_3\eta_2 \\ 
\eta_4 \\ 
1-\frac{1}{2}|\eta_2|^2-\frac{1}{2}|\eta_4|^2 \\ 
\eta_2
\end{array}\right),  \label{perturbativecp3parametrization} 
\end{equation}
where we parameterize as 
\begin{equation}
\eta_i(\boldsymbol{x})=\frac{\sigma_i(\boldsymbol{x})+i\vartheta_i(\boldsymbol{x})}{\sqrt{2}},
\label{etasigmathetaparametrization}
\end{equation}
with $i=1,2,3,4$, 
obeying the equal-time commutation relations between $\eta_i$ and $\eta_j$, or
\begin{equation}
\left[
\eta_i(\boldsymbol{x},t),\eta_j^\dagger(\boldsymbol{x},t)
\right]=\frac{2}{\rho_0}\delta_{ij}\delta(\boldsymbol{x}-\boldsymbol{y}),
\label{nu2etaccr}
\end{equation}
or
\begin{equation}
\left[
\sigma_i(\boldsymbol{x},t),\vartheta_j(\boldsymbol{x},t)
\right]=\frac{2i}{\rho_0}\delta_{ij}\delta(\boldsymbol{x}-\boldsymbol{y}). 
\label{nu2sigmathetaccr}
\end{equation}
They are required so the  SU(4) algebraic relation holds for $\mathcal{S}_a$, 
$\mathcal{P}_a$, and $\mathcal{S}_{ab}$. For a detailed discussion, see Appendix \ref{appendix}.

We calculate the isospin components \eqref{su4isospin2} with the use of $\boldsymbol{n}_{i}=U\bar{\boldsymbol{n}}_{i}$,
 and substitute them into the effective Hamiltonian \eqref{su4effectivehamiltonian1}. 
In this way we obtain the effective Hamiltonian for $\eta_i$, 
which is shown to be the same as the
one for the NG modes derived in Ref.\cite{yhama}.

\subsection{NG modes in the spin  phase}
\label{spinng}
\begin{figure}[t]
\begin{center}
\centering
\includegraphics[width=0.9\textwidth]{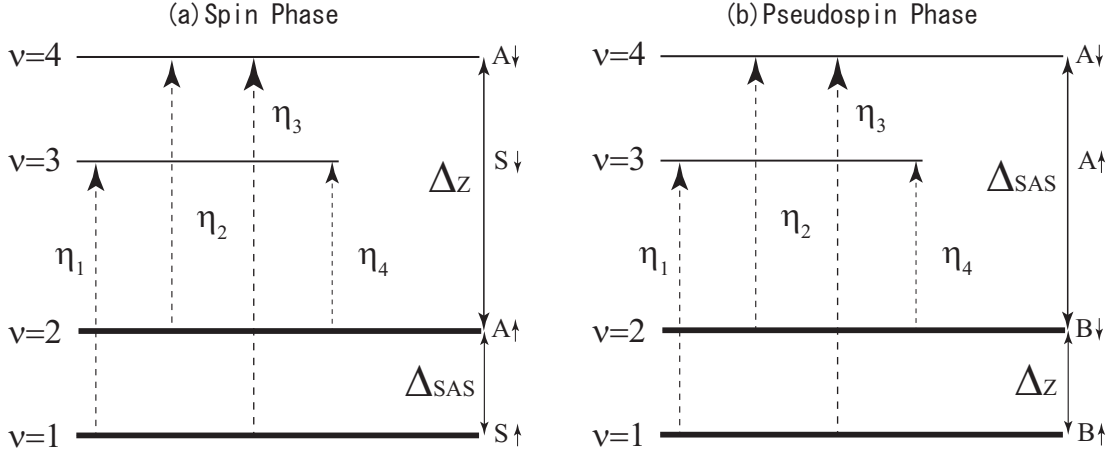} 
\end{center}
\caption{  The lowest two energy levels are occupied in the ground state at $\nu=2.$
Small fluctuations are the NG modes $\eta_1$, $\eta_2$, 
$\eta_3$, and $\eta_4$.
(a) For the spin phase, $\eta_1$ and $\eta_2$ describe the fluctuation 
from the up-spin symmetric state to the down-spin symmetric state
and from the up-spin antisymmetric state to the down-spin antisymmetric state, respectively. 
Their energy levels are  degenerated with the Zeeman gap $\Delta_{\text{Z}}$. 
On the other hand, $\eta_3$ and $\eta_4$, which are fluctuations from 
the up-spin symmetric state to the down-spin antisymmetric state
and from the up-spin antisymmetric state to the down-spin symmetric state,  
have an energy gap of $\Delta_{\text{Z}}\pm\Delta_{\text{SAS}}$, respectively. 
(b) For the pseudospin phase $\eta_1$ and $\eta_2$ describe the fluctuation 
from the up-spin bonding state to the up-spin antibonding state
and from the down-spin bonding state to the down-spin antibonding state, respectively. 
Their energy levels are  degenerated with the tunneling gap $\Delta_{\text{SAS}}$. 
On the other hand, $\eta_3$ and $\eta_4$, which are fluctuations from the
up-spin bonding state to the down-spin antibonding state
and from down-spin bonding state to the up-spin antibonding state, 
have an energy gap of $\Delta_{\text{SAS}}\pm\Delta_{\text{Z}}$, respectively.  }
\label{nu2ngspectrum} 
\end{figure}
As an illustration we study the spin phase at $\sigma_0=0$,
 where the transformation \eqref{su4rotation} is given by 
\begin{align}
U=\frac{1}{\sqrt{2}}
\left( 
\begin{array}{cccc}
1 & 0 & -1 & 0  \\
0 & 1 & 0 & -1    \\
1 & 0 & 1 & 0  \\
0 & 1 & 0 & 1    \\
\end{array} 
\right),
\end{align}
by setting $\alpha,\beta=0$. We note that 
\begin{align}
\bar{\boldsymbol{n}}= 
\left( 
\begin{array}{c}
n^{\text{S}\uparrow}   \\
n^{\text{S}\downarrow}    \\
n^{\text{A}\uparrow}  \\
n^{\text{A}\downarrow}    \\
\end{array} 
\right)=U^\dagger
\left( 
\begin{array}{c}
n^{\text{f}\uparrow}   \\
n^{\text{f}\downarrow}   \\
n^{\text{b}\uparrow}   \\
n^{\text{b}\downarrow}    \\ 
\end{array} 
\right)
=U^\dagger\boldsymbol{n},
\end{align}
where
\begin{equation}    
n^{\text{S}\alpha}=\frac{1}{\sqrt{2}}(n^{\text{b}\alpha} +n^{\text{f}\alpha} ), \quad
n^{\text{A}\alpha}=\frac{1}{\sqrt{2}}(n^{\text{b}\alpha} -n^{\text{f}\alpha} ),
\label{spingroundstate}
\end{equation}   
with $\alpha=\uparrow,\downarrow$. The lowest-energy one-body electron state  is the up-spin symmetric state,
and the second lowest energy state is the up-spin antisymmetric state. 
They are filled up at $\nu=2$. The perturbative excitations $\eta_i$ are
as illustrated in Fig. \ref{nu2ngspectrum} (a).

It follows from \eqref{su4isospin2}, \eqref{su4rotation}, and \eqref{perturbativecp3parametrization}  
that the isospin densities are explicitly given in terms of $\sigma_i(\boldsymbol{x})$ and $\vartheta_i(\boldsymbol{x})$
by
\begin{align} 
\mathcal{S}_{x}&=\frac{\sigma_1+\sigma_2}{\sqrt{2}}\equiv\tilde{\sigma}_1,\quad 
\mathcal{S}_{y}=\frac{\vartheta_1+\vartheta_2}{\sqrt{2}}\equiv\tilde{\vartheta}_1,\quad 
\mathcal{R}_{xx}=\frac{\sigma_1-\sigma_2}{\sqrt{2}}\equiv\tilde{\sigma}_2,\quad 
\mathcal{R}_{yx}=\frac{\vartheta_1-\vartheta_2}{\sqrt{2}}\equiv\tilde{\vartheta}_2,\notag\\ 
\mathcal{R}_{yy}&=\frac{\sigma_4-\sigma_3}{\sqrt{2}}\equiv-\tilde{\sigma}_3,\quad 
\mathcal{R}_{xy}=\frac{\vartheta_3-\vartheta_4}{\sqrt{2}}\equiv\tilde{\vartheta}_3,\quad
\mathcal{R}_{xz}=-\frac{\sigma_4+\sigma_3}{\sqrt{2}}\equiv\tilde{\sigma}_4,\quad 
\mathcal{R}_{yz}=-\frac{\vartheta_4+\vartheta_3}{\sqrt{2}}\equiv\tilde{\vartheta}_4,\notag\\
\mathcal{S}_{z}&=1-\sum_{i=1}^4\frac{\sigma^2_i+\vartheta^2_i}{{2}}
=1-\sum_{i=1}^4\frac{\tilde{\sigma}^2_i+\tilde{\vartheta}^2_i}{{2}},\quad
\mathcal{P}_{x}={\tilde{\sigma}_3\tilde{\sigma}_4}
+{\tilde{\vartheta}_3\tilde{\vartheta}_4},\quad
\mathcal{P}_{y}={\tilde{\sigma}_4\tilde{\vartheta}_2}
-{\tilde{\sigma}_2\tilde{\vartheta}_4},\notag\\
\mathcal{P}_{z}&=-\left({\tilde{\sigma}_2\tilde{\sigma}_3}
+{\tilde{\vartheta}_2\tilde{\vartheta}_3}\right),\quad
\mathcal{R}_{zx}=-\left({\tilde{\sigma}_1\tilde{\sigma}_2}
+{\tilde{\vartheta}_1\tilde{\vartheta}_2}\right),\quad
\mathcal{R}_{zy}={\tilde{\sigma}_3\tilde{\vartheta}_1}
-{\tilde{\sigma}_1\tilde{\vartheta}_3},\quad
\mathcal{R}_{zz}=-\left({\tilde{\sigma}_1\tilde{\sigma}_4}
+{\tilde{\vartheta}_1\tilde{\vartheta}_4}\right). 
\label{SPRgoldstonespin}
\end{align}
Substituting them into (\ref{su4effectivehamiltonian1}), we obtain the
effective Hamiltonian of the NG modes 
in terms of the canonical sets of $\tilde{\sigma}_{i }$ and $\tilde{\vartheta}_{i }$ as 
\begin{align}   
{\mathcal H}^{\text{spin}}&= {J_s}\sum_{a=1,4}\left[(\partial_k\tilde{\sigma}_{a})^2+(\partial_k\tilde{\vartheta}_{a})^2\right] 
+ J^d_s\sum_{a=2,3}\left[(\partial_k\tilde{\sigma}_{a})^2+(\partial_k\tilde{\vartheta}_{a})^2\right] \notag\\ 
&+\frac{\rho_0\Delta_{\text{Z}}}{4}\sum_{=1,4}\left[\tilde{\sigma}_{a}^2+\tilde{\vartheta}_{a}^2\right]
+\left(\frac{\rho_0\Delta_{\text{Z}}}{4}
+\rho_0\epsilon^-_X\right)\sum_{a=2,3}\left[\tilde{\sigma}_{a}^2+\tilde{\vartheta}_{a}^2\right]\notag\\
&-\frac{\rho_0\Delta_{\text{SAS}}}{2}\left[
{\tilde{\sigma}_{3}\tilde{\sigma}_{4}}+{\tilde{\vartheta}_{3}\tilde{\vartheta}_{4}}
\right]
+\frac{\rho_0\Delta_{\text{bias}}}{2}\left[{\tilde{\sigma}_{2}\tilde{\sigma}_{3}}
+{\tilde{\vartheta}_{2}\tilde{\vartheta}_{3}}
\right].
\label{spineffectivehamiltonian1}
\end{align}

The annihilation operators are defined by
\begin{align}
\eta^{\text{s}}_i(\boldsymbol{x})=\frac{{\check{\sigma}_{i}(\boldsymbol{x})+i\check{\vartheta}_{i}(\boldsymbol{x})}}
{\sqrt{2}},
\label{spincomplexfluctuation}
\end{align}
with 
\begin{equation}
\check{\sigma}_{i}\equiv\rho^{1/2}_{\Phi}\tilde{\sigma}_{i}, \qquad \check{\vartheta}_{i}\equiv\rho^{1/2}_{\Phi}\tilde{\vartheta}_{i},
\label{checksigmatheta}
\end{equation}
 and they satisfy the commutation relations  
\begin{align}
\left[ \check{\sigma}_{i}(\boldsymbol{x},t),\check{\vartheta}_j(\boldsymbol{y},t)\right]&=
i\delta_{ij}\delta(\boldsymbol{x}-\boldsymbol{y}), 
\label{checkccr}
\end{align} 
or
\begin{equation}
\left[ \eta^{\text{s}}_i(\boldsymbol{x},t),\eta_j^{\text{s}\dagger}(\boldsymbol{y},t) \right]
=\delta_{ij}\delta(\boldsymbol{x}-\boldsymbol{y}),
\end{equation} 
with $i,j=1,2,3,4$. 

The effective Hamiltonian \eqref{spineffectivehamiltonian1} reads
in terms of the creation and annihilation variables \eqref{spincomplexfluctuation} as
\begin{align}
{\mathcal H}^{\text{spin}}&=\frac{4J_s}{\rho_0}\sum_{a=1,4}\partial_k \eta_a^{\text{s}\dagger}\partial_k \eta^{\text{s}}_a
+\frac{4J^d_s}{\rho_0}\sum_{a=2,3}\partial_k \eta_a^{\text{s}\dagger}\partial_k \eta^{\text{s}}_a
+\Delta_{\text{Z}}\sum_{a=1,4}\eta_a^{\text{s}\dagger}\eta^{\text{s}}_a+[\Delta_{\text{Z}}+4\epsilon^-_X ]
\sum_{a=2,3}\eta_a^{\text{s}\dagger}\eta_a^{\text{s}}\notag\\
&+\Delta_{\text{bias}}[\eta_2^{\text{s}\dagger}\eta^{\text{s}}_3+\eta_3^{\text{s}\dagger}\eta^{\text{s}}_2]
-\Delta_{\text{SAS}}[\eta_3^{\text{s}\dagger}\eta^{\text{s}}_4+\eta_4^{\text{s}\dagger}\eta^{\text{s}}_3].
\end{align} 
The variables $\eta^{\text{s}} _{2}$, $\eta^{\text{s}} _{3}$, and $\eta^{\text{s}} _{4}$ are mixing 
by $\Delta_{\text{SAS}}$ and $\Delta_{\text{bias}}$.

In the momentum space, the annihilation and creation operators are 
$\eta^{\text{s}}_{i,\boldsymbol{k}}$ and $\eta_{i,\boldsymbol{k}}^{\text{s}\dagger}$ together with the commutation relations 
\begin{equation}
\left[ \eta^{\text{s}}_{i,\boldsymbol{k}},\eta_{j,\boldsymbol{k}^\prime}^{\text{s}\dagger}\right]=\delta_{ij}\delta(\boldsymbol{k}-\boldsymbol{k}^\prime).
\end{equation}
For the sake of the simplicity
we consider the balanced configuration with $\Delta_{\text{bias}}=0$  
in the rest of this subsection.
Then the Hamiltonian density is given by
\begin{align}
H^{\text{spin}}&=\int d^2 k \ {\mathcal H}^{\text{spin}},\notag\\
{\mathcal H}^{\text{spin}}&={\mathcal H}^{\text{spin}}_1+{\mathcal H}^{\text{spin}}_2+{\mathcal H}^{\text{spin}}_3,
\label{totalspinhamiltonian}
\end{align} 
where 
\begin{align}
{\mathcal H}^{\text{spin}}_1&=\left[ \frac{4J_s}{\rho_0}\boldsymbol{k}^2+\Delta_{\text{Z}}\right]\eta^{\text{s}\dagger}_{1,\boldsymbol{k}}\eta^{\text{s}}_{1,\boldsymbol{k}}, \label{momentumspinhamiltonian1} \\ 
{\mathcal H}^{\text{spin}}_2&=\left[ \frac{4J^d_s}{\rho_0}\boldsymbol{k}^2+\Delta_{\text{Z}}+4\epsilon^-_X \right]\eta_{2,\boldsymbol{k}}^{\text{s}\dagger}\eta^{\text{s}}_{2,\boldsymbol{k}}, \label{momentumspinhamiltonian2} \\
{\mathcal H}^{\text{spin}}_3&=\left[ \frac{4J^d_s}{\rho_0}\boldsymbol{k}^2+\Delta_{\text{Z}}+4\epsilon^-_X\right]\eta_{3,\boldsymbol{k}}^{\text{s}\dagger}\eta^{\text{s}}_{3,\boldsymbol{k}}
+\left[ \frac{4J_s}{\rho_0}\boldsymbol{k}^2+\Delta_{\text{Z}}\right]
 \eta_{4,\boldsymbol{k}}^{\text{s}\dagger}\eta^{\text{s}}_{4,\boldsymbol{k}} 
-\Delta_{\text{SAS}}\left[\eta_{3,\boldsymbol{k}}^{\text{s}\dagger}\eta^{\text{s}}_{4,\boldsymbol{k}}+\eta_{4,\boldsymbol{k}}^{\text{s}\dagger}\eta_{3,\boldsymbol{k}}^{\text{s}} \right].
\label{momentumspinhamiltonian3} 
\end{align}

We first analyze the dispersion relation and the coherence length of $ \eta^{\text{s}}_{1,\boldsymbol{k}}$. 
From \eqref{momentumspinhamiltonian1}, we have 
\begin{align}
E_{\eta^{\text{s}}_1}(\boldsymbol{k})&=\frac{4J_s}{\rho_0}\boldsymbol{k}^2+\Delta_{\text{Z}}, \label{spindispersion1} \\ 
\xi_{\eta^{\text{s}}_1}&=2l_B\sqrt{\frac{\pi J_s}{\Delta_{\text{Z}}}}. \label{spincoherencelength1}
\end{align}
The coherent length diverges in the limit $\Delta _{\text{Z}}\rightarrow 0$.
This mode is a pure spin wave since it describes the fluctuation of $\mathcal{S}_x$ and $\mathcal{S}_y$ as in \eqref{SPRgoldstonespin}. 
Indeed, the energy \eqref{spindispersion1}, as well as the coherent length (\ref{spincoherencelength1}),
depend only on the Zeeman gap $\Delta_{\text{Z}}$ and the intralayer stiffness $J_s$.   

We next analyze those of $\eta^{\text{s}}_{2,\boldsymbol{k}}$:
\begin{align}
E_{\eta^{\text{s}}_2}(\boldsymbol{k})&=\frac{4J^d_s}{\rho_0}\boldsymbol{k}^2+\Delta_{\text{Z}}+4\epsilon^-_X,  \label{spindispersion2} \\
\xi_{\eta^{\text{s}}_2}&=2l_B\sqrt{\frac{\pi J^d_s}{\Delta_{\text{Z}}+4\epsilon^-_X}}. \label{spincoherencelength2}
\end{align} 
They depend not only on $\Delta_{\text{Z}}$ but also on the exchange Coulomb energy $\epsilon_X^-$
and the interlayer stiffness originating in the interlayer Coulomb interaction. 
This mode is a $R$-spin wave since it describes the fluctuation of $\mathcal{R}_{xx}$ and $\mathcal{R}_{yx}$. 
From \eqref{spindispersion1} and \eqref{spindispersion2} we see that, in the one body picture, 
$\eta^{\text{s}}_1$ and $\eta^{\text{s}}_2$ have the same energy gap $\Delta_{\text{Z}}$. 
Indeed, they are described in terms of $\eta_1$ and $\eta_2$, having the same energy gap $\Delta_{\text{Z}}$ (Fig. \ref{nu2ngspectrum} (a)).

We finally analyze those of 
$\eta^{\text{s}}_{3,\boldsymbol{k}}$ and $\eta^{\text{s}}_{4,\boldsymbol{k}}$,
which are coupled. Hamiltonian \eqref{momentumspinhamiltonian3} can be written, in  matrix form, 
\begin{align} 
{\mathcal H}^{\text{spin}}_3=\left( 
\begin{array}{c}
\eta^{\text{s}}_{3,\boldsymbol{k}}   \\
\eta^{\text{s}}_{4,\boldsymbol{k}}  \\
\end{array} 
\right)^\dagger
\left( 
\begin{array}{cc}
A_{\boldsymbol{k}} & -\Delta_{\text{SAS}}  \\
-\Delta_{\text{SAS}} & B_{\boldsymbol{k}} \\
\end{array} 
\right)
\left( 
\begin{array}{c}
\eta^{\text{s}}_{3,\boldsymbol{k}}   \\
\eta^{\text{s}}_{4,\boldsymbol{k}}  \\
\end{array} 
\right), 
\label{momentumspinmatrixhamiltonian}
\end{align} 
where
\begin{align}
A_{\boldsymbol{k}}=\frac{4J^d_s}{\rho_0}\boldsymbol{k}^2+\Delta_{\text{Z}}+4\epsilon^-_X, \quad
B_{\boldsymbol{k}}=\frac{4J_s}{\rho_0}\boldsymbol{k}^2+\Delta_{\text{Z}}.
\label{spinAandB}
\end{align}
Hamiltonian \eqref{momentumspinmatrixhamiltonian} can be diagonalized as 
\begin{align} 
{\mathcal H}^{\text{spin}}_3=\left( 
\begin{array}{c}
\tilde{\eta}^{\text{s}}_{3,\boldsymbol{k}}   \\
\tilde{\eta}^{\text{s}}_{4,\boldsymbol{k}}  \\
\end{array} 
\right)^\dagger
\left( 
\begin{array}{cc}
E^{\tilde{\eta}^{\text{s}}_3} & 0  \\
0 & E^{\tilde{\eta}^{\text{s}}_4} \\
\end{array} 
\right)
\left( 
\begin{array}{c}
\tilde{\eta}^{\text{s}}_{3,\boldsymbol{k}}   \\
\tilde{\eta}^{\text{s}}_{4,\boldsymbol{k}}  \\  
\end{array} 
\right),
\label{newspinmixinghamiltonian}
\end{align}
where 
\begin{align}
E^{\tilde{\eta}^{\text{s}}_3}&=\frac{1}{2}
\left[A_{\boldsymbol{k}}+B_{\boldsymbol{k}}+
\sqrt{(A_{\boldsymbol{k}}-B_{\boldsymbol{k}})^2+4\Delta^2_{\text{SAS}}} \right],\quad
E^{\tilde{\eta}^{\text{s}}_4}=\frac{1}{2}
\left[A_{\boldsymbol{k}}+B_{\boldsymbol{k}}-
\sqrt{(A_{\boldsymbol{k}}-B_{\boldsymbol{k}})^2+4\Delta^2_{\text{SAS}}} \right],
\label{spindispersion3}
\end{align}
and the annihilation operator $\tilde{\eta}^{\text{s}}_{i,\boldsymbol{k}}$ ($i=3,4$) given by the form
\begin{align}
\tilde{\eta}^{\text{s}}_{3,\boldsymbol{k}}&=
\frac{\left(\sqrt{C^{2}_{\boldsymbol{k}}+4\Delta^2_{\text{SAS}}}+C_{\boldsymbol{k}}\right)\eta_{3,\boldsymbol{k}}
-2\Delta_{\text{SAS}}\eta_{4,\boldsymbol{k}}}
{\sqrt{2\left(C^{2}_{\boldsymbol{k}}+4\Delta^2_{\text{SAS}}+C_{\boldsymbol{k}} \sqrt{C^{2}_{\boldsymbol{k}}+4\Delta^2_{\text{SAS}}}
\right)}},\notag\\
\tilde{\eta}^{\text{s}}_{4,\boldsymbol{k}}&=
\frac{\left(\sqrt{C^{2}_{\boldsymbol{k}}+4\Delta^2_{\text{SAS}}}-C_{\boldsymbol{k}}\right)\eta_{3,\boldsymbol{k}}
+2\Delta_{\text{SAS}}\eta_{4,\boldsymbol{k}}}
{\sqrt{2\left(C^{2}_{\boldsymbol{k}}+4\Delta^2_{\text{SAS}}-C_{\boldsymbol{k}} \sqrt{C^{2}_{\boldsymbol{k}}+4\Delta^2_{\text{SAS}}}
\right)}},\label{newspinannihilation}
\end{align} 
with $C_{\boldsymbol{k}}=A_{\boldsymbol{k}}-B_{\boldsymbol{k}}$. 
The annihilation operators \eqref{newspinannihilation} 
satisfy the commutation relations
\begin{align}
\left[\tilde{\eta}^{\text{s}}_{i,\boldsymbol{k}},\tilde{\eta}^{\text{s}\dagger}_{j,\boldsymbol{k}^\prime}\right]
=\delta_{ij}\delta(\boldsymbol{k}-\boldsymbol{k}^\prime), 
\end{align}
with $i,j=3,4$. 
We obtain the dispersions for the modes  $\tilde{\eta}^{\text{s}}_{i,\boldsymbol{k}}$ ($i=3,4$) 
from \eqref{spinAandB} and \eqref{spindispersion3}.

By taking the limit $\boldsymbol{k}\rightarrow0$ in \eqref{spindispersion3},
we have two gaps
\begin{align}
E^{\tilde{\eta}^{\text{s}}_3}_{\boldsymbol{k}=0}&=
\Delta_{\text{Z}}+2\epsilon^-_X+\left[4(\epsilon^-_X)^2+\Delta^2_{\text{SAS}}\right]^{\frac{1}{2}},\quad
E^{\tilde{\eta}^{\text{s}}_4}_{\boldsymbol{k}=0}=
\Delta_{\text{Z}}+2\epsilon^-_X-\left[4(\epsilon^-_X)^2+\Delta^2_{\text{SAS}}\right]^{\frac{1}{2}}.
\label{spingaps}
\end{align}
The gapless condition $(E^{\tilde{\eta}^{\text{s}}_4}_{\boldsymbol{k}=0}=0)$ implies
\begin{equation}
\Delta_{\text{Z}}(\Delta_{\text{Z}}+4\epsilon^-_X)-\Delta_{\text{SAS}}^2=0,
\end{equation}
which holds only along the boundary of the spin and CAF phases: see (4.17) in Ref.\cite{Ezawa:2005xi}.
In the interior of the spin phase we have $\Delta_{\text{Z}}(\Delta_{\text{Z}}+4\epsilon^-_X)-\Delta_{\text{SAS}}^2>0$,
which implies that   no gapless modes arise from 
$\tilde{\eta}^{\text{s}}_3$ and $\tilde{\eta}^{\text{s}}_4$.
From \eqref{spingaps}, in the one body picture, 
$\tilde{\eta}^{\text{s}}_3$ and $\tilde{\eta}^{\text{s}}_4$ 
have the energy gap $\Delta_{\text{Z}}\pm\Delta_{\text{SAS}}$, respectively. 
Indeed they are described in terms of $\eta_3$ and $\eta_4$ (Fig. \ref{nu2ngspectrum} (a)).  
These excitation modes are $R$-spin waves coupled with the layer degree of freedom. 
There emerge four complex NG modes, one  describing the spin wave ($\eta^{\text{s}}_1$), and the other three the $R$-spin waves 
($\eta^{\text{s}}_2,\eta^{\text{s}}_3,\eta^{\text{s}}_4$).   

\subsection{NG modes in the  pseudospin phase}
\label{ppinng}
For the pseudospin phase,  $\beta$ is identified with the imbalanced parameter $\sigma_0$,  
as we  discussed in Sect. \ref{nu2groundstatestructure} with \eqref{ppinorderparameters1}.
In this subsection, instead of $\beta$ we express the effective Hamiltonian, the dispersions, and the coherence length 
in terms of $\sigma_0$, since it is a physical variable. 

From \eqref{su4rotation}, by setting $\alpha=1$, we have   
\begin{align}
U=\frac{1}{\sqrt{2}}
\left( 
\begin{array}{cccc}
\sqrt{1+\sigma_0} & -\sqrt{1-\sigma_0} & 0 & 0  \\
0 & 0 & -\sqrt{1+\sigma_0} & -\sqrt{1-\sigma_0}    \\
\sqrt{1-\sigma_0} & \sqrt{1+\sigma_0} & 0 & 0  \\
0 & 0 & -\sqrt{1-\sigma_0} & \sqrt{1+\sigma_0}    \\
\end{array} 
\right), 
\end{align}
and
\begin{align}
\bar{\boldsymbol{n}}=
\left( 
\begin{array}{c}
n^{\text{B}\uparrow}   \\
n^{\text{A}\uparrow}    \\
-n^{\text{B}\downarrow}  \\
n^{\text{A}\downarrow}    \\
\end{array} 
\right)=U^\dagger
\left( 
\begin{array}{c}
n^{\text{f}\uparrow}   \\
n^{\text{f}\downarrow}   \\
n^{\text{b}\uparrow}   \\
n^{\text{b}\downarrow}    \\ 
\end{array} 
\right)
=U^\dagger\boldsymbol{n},
\end{align} 
where
\begin{equation}    
n^{\text{B}\alpha}=\frac{1}{\sqrt{2}}(\sqrt{1-\sigma_0} n^{\text{b}\alpha} +\sqrt{1+\sigma_0} n^{\text{f}\alpha} ), \quad
n^{\text{A}\alpha}=\frac{1}{\sqrt{2}}(\sqrt{1+\sigma_0} n^{\text{b}\alpha} -\sqrt{1-\sigma_0} n^{\text{f}\alpha} ),
\label{bondingstate}
\end{equation}
with $\alpha=\uparrow,\downarrow$. 
The lowest-energy one-body electron state  is the up-spin bonding state,
and the second lowest energy state is the down-spin bonding state. 
They are filled up at $\nu=2$. The perturbative excitations $\eta_i$ 
are as illustrated in Fig. \ref{nu2ngspectrum} (b).

We go on to derive the effective Hamiltonian governing these NG
modes. 
From \eqref{su4isospin2}, \eqref{su4rotation}, and \eqref{perturbativecp3parametrization},  
the isospin densities are given in terms of $\tilde{\sigma}_i(\boldsymbol{x})$ and $\tilde{\vartheta}_i(\boldsymbol{x})$ as:
\begin{align}
\mathcal{P}_{x}& =\sigma_0\tilde{\sigma}_{2}+\sqrt{1-\sigma_0^2}\left(
1-\sum_{i=1}^4\frac{\tilde{\sigma}^2_i+\tilde{\vartheta}^2_i}{{2}}
\right), \quad
\quad
\mathcal{P}_{z} =-\sqrt{1-\sigma_0^2}\tilde{\sigma}_{2}+\sigma_0\left(
1-\sum_{i=1}^4\frac{\tilde{\sigma}^2_i+\tilde{\vartheta}^2_i}{{2}}
\right),
\notag \\
\mathcal{S}_{x}& =-\left({\tilde{\sigma}_{1}\tilde{\sigma}_{4}}+{\tilde{\vartheta}_{1}\tilde{\vartheta}_{4}}\right),\quad 
\mathcal{S}_{y} ={\tilde{\sigma}_{1}\tilde{\vartheta}_{3}}-{\tilde{\sigma}_{3}\tilde{\vartheta}_{1}},\quad
\mathcal{S}_{z} ={\tilde{\sigma}_{3}\tilde{\sigma}_{4}}+{\tilde{\vartheta}_{3}\tilde{\vartheta}_{4}},\notag\\
\mathcal{R}_{zy}& =\tilde{\vartheta}_{1},\qquad
\mathcal{P}_{y} =\tilde{\vartheta}_{2}, \qquad
\mathcal{R}_{xy} =\tilde{\vartheta}_{3}, \qquad
\mathcal{R}_{yy} =\tilde{\sigma}_{4},\notag\\  
\mathcal{R}_{xx}& =
-\sqrt{1-\sigma_0^2}\left({\tilde{\sigma}_{2}\tilde{\sigma}_{3}}
+{\tilde{\vartheta}_{2}\tilde{\vartheta}_{3}}\right)
+\sigma_0\tilde{\sigma}_{3},\ 
\mathcal{R}_{xz} =
-\sigma_0\left({\tilde{\sigma}_{2}\tilde{\sigma}_{3}}+
{\tilde{\vartheta}_{2}\tilde{\vartheta}_{3}}\right)
-\sqrt{1-\sigma_0^2}\tilde{\sigma}_{3},\notag\\ 
\mathcal{R}_{yx}&=
\sqrt{1-\sigma_0^2}\left({\tilde{\sigma}_{2}\tilde{\vartheta}_{4}}
-{\tilde{\sigma}_{4}\tilde{\vartheta}_{2}}\right)
-\sigma_0\tilde{\vartheta}_{4},\
\mathcal{R}_{yz}=
\sigma_0\left({\tilde{\sigma}_{2}\tilde{\vartheta}_{4}}-{\tilde{\sigma}_{4}\tilde{\vartheta}_{2}}\right)
+\sqrt{1-\sigma_0^2}\tilde{\vartheta}_{4},\notag\\
\mathcal{R}_{zx}& =
-\sqrt{1-\sigma_0^2}\left({\tilde{\sigma}_{1}\tilde{\sigma}_{2}}+
{\tilde{\vartheta}_{1}\tilde{\vartheta}_{2}}\right)
+\sigma_0\tilde{\sigma}_{1},\
\mathcal{R}_{zz} =
-\sigma_0\left({\tilde{\sigma}_{1}\tilde{\sigma}_{2}}
+{\tilde{\vartheta}_{1}\tilde{\vartheta}_{2}}\right)
-\sqrt{1-\sigma_0^2}\tilde{\sigma}_{1}
\label{ppingoldstoneexpansion}
\end{align}
Now,  we substitute the isospin densities \eqref{ppingoldstoneexpansion} 
into the effective Hamiltonian (\ref{su4effectivehamiltonian1}).
In this way we derive the effective Hamiltonian of the NG modes in
terms of the canonical sets of $\tilde{\sigma}_{i}$ and $\tilde{\vartheta}_{i}$
(or with $\check{\sigma}_{i}$ and $\check{\vartheta}_{i}$). 

In the momentum space, this reads
\begin{align}
\int d^2 k{\mathcal H}^{\text{p}}=\int d^2 k{\mathcal H}^{\text{p}}_1
+\int d^2 k{\mathcal H}^{\text{p}}_2+\int d^2 k{\mathcal H}^{\text{p}}_3,
\label{totalppinhamiltonian}
\end{align}
where 
\begin{align}
{\mathcal H}^{\text{p}}_1&=A^{\text{p}}_{\boldsymbol{k}}\check{\sigma}^{\dagger}_{1,\boldsymbol{k}}\check{\sigma}_{1,\boldsymbol{k}}
+B^{\text{p}}_{\boldsymbol{k}}\check{\vartheta}_{1,\boldsymbol{k}}^{\dagger}\check{\vartheta}_{1,\boldsymbol{k}}, 
 \label{momentumppinhamiltonian1} \\   
{\mathcal H}^{\text{p}}_2&= C^{\text{p}}_{\boldsymbol{k}}\check{\sigma}_{2,\boldsymbol{k}}^{\dagger}\check{\sigma}_{2,\boldsymbol{k}}
+B^{\text{p}}_{\boldsymbol{k}}\check{\vartheta}_{2,\boldsymbol{k}}^{\dagger}\check{\vartheta}_{2,\boldsymbol{k}},
\label{momentumppinhamiltonian2} \\    
{\mathcal H}^{\text{p}}_3&=
(\vec{P}^{\text{p}}_{\boldsymbol{k}})^\dagger 
\mathcal{M}^{\text{p}} 
\vec{P}^{\text{p}}_{\boldsymbol{k}},
\label{momentumppinhamiltonian3}    
\end{align}
with $\check{\sigma}_{i,\boldsymbol{k}}$, and 
$\check{\vartheta}_{i,\boldsymbol{k}}$ given by \eqref{checksigmatheta}, and   
\begin{align}
A^{\text{p}}_{\boldsymbol{k}}&=\frac{2J_1^{\sigma_0}}{\rho_0}k^2+\frac{\Delta_{\text{SAS}}}{2\sqrt{1-\sigma_0^2}}-2\epsilon^-_X(1-\sigma_0^2),\quad
B^{\text{p}}_{\boldsymbol{k}}=\frac{2J^d_s}{\rho_0}k^2+\frac{\Delta_{\text{SAS}}}{2\sqrt{1-\sigma_0^2}},\notag\\
C^{\text{p}}_{\boldsymbol{k}}&=\frac{2J_1^{\sigma_0}}{\rho_0}k^2+\frac{\Delta_{\text{SAS}}}{2\sqrt{1-\sigma_0^2}}
+\epsilon_{\text{cap}}(1-\sigma_0^2),\quad
J_1^{\sigma_0}=(1-\sigma_0^2)J_s+\sigma_0^2 J^d_s,\notag\\
\vec{P}^{ \text{p}}_{\boldsymbol{k}}&= 
\left(\begin{array}{c}
\check{\vartheta}_4 \\ 
\check{\vartheta}_3 \\ 
\check{\sigma}_3 \\ 
\check{\sigma}_4
\end{array}\right),
\quad
\mathcal{M}^{\text{p}} = 
\left(  
\begin{array}{cccc} 
A^{\text{p}}_{\boldsymbol{k}} & -\Delta_{\text{Z}}/2 & 0 & 0   \\
-\Delta_{\text{Z}}/2 & B^{\text{p}}_{\boldsymbol{k}} & 0 & 0   \\ 
0 & 0 & A^{\text{p}}_{\boldsymbol{k}} & -\Delta_{\text{Z}}/2     \\
0 & 0 & -\Delta_{\text{Z}}/2 & B^{\text{p}}_{\boldsymbol{k}}     \\ 
\end{array} 
\right).
\label{ppinvariable}
\end{align}
 
We first analyze the dispersions  and the coherence lengths from
 \eqref{momentumppinhamiltonian2}, since it describes the pseudospin wave. 
It is diagonalized as: 
\begin{equation}
H^\text{p}_{2}=\int d^2 k E^\text{p}_{2}\eta^{\text{p}\dagger}_{2,\boldsymbol{k}}\eta^\text{p}_{2,\boldsymbol{k}}
\end{equation}
with 
\begin{align}
E^\text{p}_{2,\boldsymbol{k}}&=2\sqrt{B^{\text{p}}_{\boldsymbol{k}}C^{\text{p}}_{\boldsymbol{k}}},\label{etap1dispersion}\\
\eta^\text{p}_{2,\boldsymbol{k}}&=
\frac{1}{\sqrt{2}}
\left(\left(\frac{C^{\text{p}}_{\boldsymbol{k}} }{B^{\text{p}}_{\boldsymbol{k}} }\right)^{\frac{1}{4}}\check{\sigma}_{2,\boldsymbol{k}}
+i\left(\frac{B^{\text{p}}_{\boldsymbol{k}} }{C^{\text{p}}_{\boldsymbol{k}} }\right)^{\frac{1}{4}}\check{\vartheta}_{2,\boldsymbol{k}}\right),
\end{align}
where $\eta^\text{p}_{2,\boldsymbol{k}}$ satisfy the commutation relation 
\begin{equation}
\left[ \eta^{\text{p}}_{2,\boldsymbol{k}},\eta_{2,\boldsymbol{k}^\prime}^{\text{p}\dagger} \right]
=\delta(\boldsymbol{k}-\boldsymbol{k}^\prime). 
\end{equation}

Since  the ground state is a squeezed coherent state due to the capacitance energy $\epsilon_{\text{cap}}$, it is more convenient\cite{Ezawa:2008ae} to use the dispersion and the coherence lengths of 
$\check{\sigma}_{2}$ and $\check{\vartheta}_{2}$ separately.
The dispersion relations are given by
\begin{align}
E_{\boldsymbol{k}}^{\check{\sigma}_{2}}=\frac{2J^{\sigma_0}_1}{\rho_0}\boldsymbol{k}^2+\frac{\Delta_{\text{SAS}}}{2\sqrt{1-\sigma_0^2}}
+\epsilon_{\text{cap}}(1-\sigma_0^2),\quad
E_{\boldsymbol{k}}^{\check{\vartheta}_{2}}=\frac{2J_s^d}{\rho_0}\boldsymbol{k}^2+\frac{\Delta_{\text{SAS}}}{2\sqrt{1-\sigma_0^2}},
\label{ppindispersions1}
\end{align}
and their coherence lengths are
\begin{align}
\xi^{\check{\sigma}_{2}}=2l_B\sqrt{\frac{\pi J^{\sigma_0}_1}{\frac{\Delta_{\text{SAS}}}{\sqrt{1-\sigma_0^2}}
+ 2\epsilon_{\text{cap}}(1-\sigma_0^2)}}, \quad
\xi^{\check{\vartheta}_{2}}=2l_B\sqrt{\frac{\pi J^d_s\sqrt{1-\sigma_0^2}}{\Delta_{\text{SAS}}}}.
\label{ppincoherencelengths1}    
\end{align}

A similar analysis can be adopted for \eqref{momentumppinhamiltonian1},
which is diagonalized as: 
\begin{equation}
H^\text{p}_{1}=\int d^2 k E^\text{p}_{1}\eta^{\text{p}\dagger}_{1,\boldsymbol{k}}\eta^\text{p}_{1,\boldsymbol{k}}
\end{equation}
with
\begin{align}
E^\text{p}_{1}&=2\sqrt{B^{\text{p}}_{\boldsymbol{k}}A^{\text{p}}_{\boldsymbol{k}}},\label{etap2dispersion}\\
\eta^\text{p}_{1,\boldsymbol{k}}&= 
\frac{1}{\sqrt{2}}
\left(\left(\frac{A^{\text{p}}_{\boldsymbol{k}} }{B^{\text{p}}_{\boldsymbol{k}} }\right)^{\frac{1}{4}}\check{\sigma}_{1,\boldsymbol{k}}
+i\left(\frac{B^{\text{p}}_{\boldsymbol{k}} }{A^{\text{p}}_{\boldsymbol{k}} }\right)^{\frac{1}{4}}\check{\vartheta}_{1,\boldsymbol{k}}\right),
\end{align}
where $\eta^\text{p}_{1,\boldsymbol{k}}$ satisfy the commutation relation 
\begin{equation}
\left[ \eta^{\text{p}}_{1,\boldsymbol{k}},\eta_{1,\boldsymbol{k}^\prime}^{\text{p}\dagger} \right]
=\delta(\boldsymbol{k}-\boldsymbol{k}^\prime). 
\end{equation} 
 
The dispersion relations of the canonical sets of 
$\check{\sigma}_{1}$ and $\check{\vartheta}_{1}$ are given by 
\begin{align}
E_{\boldsymbol{k}}^{\check{\sigma}_{1}}=\frac{2J^{\sigma_0}_1}{\rho_0}\boldsymbol{k}^2+\frac{\Delta_{\text{SAS}}}{2\sqrt{1-\sigma_0^2}}
-2\epsilon^-_X(1-\sigma_0^2),\quad
E_{\boldsymbol{k}}^{\check{\vartheta}_{1}}=\frac{2J_s^d}{\rho_0}\boldsymbol{k}^2+\frac{\Delta_{\text{SAS}}}{2\sqrt{1-\sigma_0^2}}.
\label{ppindispersions2}
\end{align}
Their coherence lengths are
\begin{align}
\xi^{\check{\sigma}_{1}}&=2l_B\sqrt{\frac{\pi J^{\sigma_0}_1}{\frac{\Delta_{\text{SAS}}}{\sqrt{1-\sigma_0^2}}
-4\epsilon_X^-(1-\sigma_0^2)}}, \quad 
\xi^{\check{\vartheta}_{1}}=2l_B\sqrt{\frac{\pi J^d_s\sqrt{1-\sigma_0^2}}{\Delta_{\text{SAS}}}}
\label{StepA}.  
\end{align}  
It appears that $\xi ^{\check{\sigma}_{1}}$ is ill-defined for $%
\Delta _{\text{SAS}}\rightarrow 0$ in (\ref{StepA}). This is not the case
due to the relation (\ref{StepB}) in the pseudospin phase, which we mention 
soon.
We see that from \eqref{etap1dispersion} and \eqref{etap2dispersion}, in the one body picture, 
$\eta^{\text{p}}_1$ and $\eta^{\text{p}}_2$ have the same energy gap $\Delta_{\text{SAS}}$. 
They are described in terms of $\eta_1$ and $\eta_2$, having the same energy gap $\Delta_{\text{SAS}}$ (Fig. \ref{nu2ngspectrum} (b)).

Finally,   analyzing of  the Hamiltonian \eqref{momentumppinhamiltonian3}
as in the case of the spin phase, 
we obtain the condition for the  
existence of a gapless mode:
 \begin{equation}
\frac{\Delta_{\text{SAS}}}{\sqrt{1-\sigma_0^2}}
\left[ 
\frac{\Delta_{\text{SAS}}}{\sqrt{1-\sigma_0^2}}-4\epsilon^-_X(1-\sigma_0^2)
\right]-\Delta^2_{\text{Z}}=0. 
\end{equation}
This occurs along the pseudospin-canted boundary: see (5.3) and (5.4) in Ref. \cite{Ezawa:2005xi}. 
Inside the pseudospin phase, since we have
\begin{equation}
\frac{\Delta_{\text{SAS}}}{\sqrt{1-\sigma_0^2}}
\left[ 
\frac{\Delta_{\text{SAS}}}{\sqrt{1-\sigma_0^2}}-4\epsilon^-_X(1-\sigma_0^2)  
\right]-\Delta^2_{\text{Z}}>0, \label{StepB}
\end{equation}
there are no gapless modes.  
 
\subsection{NG modes in the CAF phase} 
\label{effectivehamiltoniancaf}
We derive the effective Hamiltonian of the NG modes in terms of the canonical sets of
$\check{\sigma}_{i }$ and 
$\check{\vartheta}_{i }$. 
This can be done by substituting \eqref{cafisospinrelation1} and \eqref{cafisospinrelation2}
into the Hamiltonian \eqref{su4effectivehamiltonian1}. 
We first derive the Hamiltonian, without taking any limits.  
Since the expression becomes too extensive, we introduce the notation
\begin{align}
c_{\theta_\alpha} &\equiv \cos \theta_\alpha, \quad
s_{\theta_\alpha} \equiv \sin {\theta_\alpha}, \quad
c_{\theta_\beta} \equiv \cos {\theta_\beta}, \quad
s_{\theta_\beta} \equiv \sin {\theta_\beta}, \quad
c_{\theta_\delta}\equiv \cos \theta_\delta, \quad
s_{\theta_\delta} \equiv \sin \theta_\delta.
\label{sincosdefinition2} 
\end{align}
to make the expression for  the effective Hamiltonian more manageable. 

Working in the momentum space,  the effective Hamiltonian reads 
\begin{align}
H^{\text{c}}=\int d^2 k{\mathcal{H}}^{\text{c}}=
\int d^2 k{\mathcal{H}}^{\text{c}}_{1}+
\int d^2 k{\mathcal{H}}^{\text{c}}_{2}, 
\end{align}
where
\begin{align}  
{\mathcal{H}}^{\text{c}}_{1}&=      
\left(
\frac{2}{\rho_0} J^{{\alpha}}_1 \boldsymbol{k}^2+\frac{\Delta_0 c^{-1}_{\theta_{\beta}}}{2}
\right)
\check{\vartheta}^{\dagger}_{1,\boldsymbol{k}} \check{\vartheta}_{1,\boldsymbol{k}}+
\left(
\frac{2}{\rho_0}(c_{\theta_\delta}^2 J_s+s_{\theta_\delta}^2 J_1^\beta) \boldsymbol{k}^2+\frac{M-4 (s^2_{\theta_\delta} c_{\theta_\beta}^2
 +c_{\theta_\delta}^2)\epsilon^{-}_X}{2} 
\right)
\check{\sigma}^{\dagger}_{1,\boldsymbol{k}} \check{\sigma}_{1,\boldsymbol{k}},
\label{scantedhamiltonian1}\\
{\mathcal{H}}^{\text{c}}_{2}&=\vec{Q}^{\text{c}\dagger}_{\boldsymbol{k}} 
{\mathcal{M}}^{\text{c}}_2\vec{Q}^{\text{c}}_{\boldsymbol{k}},
\label{scantedhamiltonian2}
\end{align} 
with   
\begin{align}
J^{{\alpha}}_1&=c_{\theta_\alpha}^2 J_s+s_{\theta_\alpha}^2 J^d_s, \quad
M=4c_{\theta_\alpha}^2\epsilon^-_X +\Delta_0 c_{\theta_\beta}^{-1},\notag\\
\vec{Q}^{\text{c}}_{\boldsymbol{k}}&=  
\left(
\begin{array}{c}
\check{\vartheta}_{2,\boldsymbol{k}}   \\
\check{\vartheta}_{4,\boldsymbol{k}}  \\
\check{\vartheta}_{3,\boldsymbol{k}}   \\ 
\check{\sigma}_{2,\boldsymbol{k}}   \\
\check{\sigma}_{4,\boldsymbol{k}}   \\
\check{\sigma}_{3,\boldsymbol{k}}   \\
\end{array}
\right), \quad
{\mathcal{M}}_2^{\text{c}}=  
\left(
\begin{array}{cccccc}
A^{\text{c}} &  c^{\text{c}}  & -e^{\text{c}} & 0 & 0 & 0   \\
c^{\text{c}} & C^{\text{c}} & -f^{\text{c}} & 0 & 0 & 0 \\
-e^{\text{c}}  & -f^{\text{c}}  & F^{\text{c}} & 0 & 0 & 0  \\ 
0 & 0 & 0 & B^{\text{c}} &  a^{\text{c}}  & b^{\text{c}}  \\
0 & 0 & 0 & a^{\text{c}} & D^{\text{c}} & d^{\text{c}}  \\
0 & 0 & 0 & b^{\text{c}}  & d^{\text{c}}  & E^{\text{c}}  \\
\end{array}
\right).
\label{matrixscaf}
\end{align}
The Matrix elements in \eqref{matrixscaf}  are given by 
\begin{align} 
A^{\text{c}}&=\frac{2\boldsymbol{k}^2}{\rho_0}\left[c_{\theta_\delta}^2 J_3^\beta+s_{\theta_\delta}^2 J^d_s\right]
+\frac{M}{2}-2 s_{\theta_\beta}^2 c_{\theta_\delta}^2\epsilon_X^-, \quad
B^{\text{c}}=\frac{2\boldsymbol{k}^2}{\rho_0}\left[c_{\theta_\alpha}^2 J_3^\beta +s_{\theta_\alpha}^2 J_1^\beta\right]+\frac{\Delta_0 }{2c_{\theta_\beta}}
+\frac{c_{\theta_\beta}^2\epsilon_\alpha }{2}, \notag\\
C^{\text{c}}&=\frac{2\boldsymbol{k}^2}{\rho_0} J^\beta_1 +\frac{M}{2}-2 c_{\theta_\beta}^2\epsilon_X^-,\quad 
D^{\text{c}}=\frac{2\boldsymbol{k}^2}{\rho_0}\left[c_{\theta_\delta}^2 \left(s_{\theta_\alpha}^2 J_3^\beta+c_{\theta_\alpha}^2 J_1^\beta\right)+s_{\theta_\delta}^2 J^{\alpha}_1\right]
+\frac{\Delta_0 }{2c_{\theta_\beta}}
+\frac{c_{\theta_\delta}^2 s_{\theta_\beta}^2\epsilon_\alpha }{2}, \notag\\
E^{\text{c}}&=\frac{2\boldsymbol{k}^2}{\rho_0}\left[s_{\theta_\delta}^2 \left(c_{\theta_\alpha}^2 J^\beta_3+s_{\theta_\alpha}^2 J^\beta_1\right)+c_{\theta_\delta}^2J^\alpha_3 \right] 
+\frac{M}{2}+s_{\theta_\beta}^2 s_{\theta_\delta}^2 c_{\theta_\alpha}^2\epsilon_{\text{cap}}
-2 (c_{\theta_\beta}^2 s_{\theta_\delta}^2+c_{\theta_\delta}^2) s_{\theta_\alpha}^2\epsilon_X^-, \notag\\
F^{\text{c}}&=\frac{2\boldsymbol{k}^2}{\rho_0} J^d_s +\frac{M}{2},
\label{matrixelements1}
\end{align}
and   
\begin{align}
a^{\text{c}}&=\frac{2\boldsymbol{k}^2}{\rho_0}c_{\theta_\delta}c_{2\theta_\alpha} J_2^{\beta}+\frac{s_{2\theta_\beta}  c_{\theta_\delta}}{4} \epsilon_{\alpha}, \quad  
b^{\text{c}}=-\frac{2\boldsymbol{k}^2}{\rho_0}s_{\theta_\delta} s_{2\theta_\alpha}  J_2^\beta +L+\frac{\Delta_{\text{SAS}}}{4\Delta_0}c_{\theta_\alpha} s_{2\theta_\beta} \epsilon_{\alpha},\notag\\
c^{\text{c}}&=\frac{2\boldsymbol{k}^2}{\rho_0}c_{\theta_\delta} J_2^\beta  + s_{2\theta_\beta}  c_{\theta_\delta}\epsilon_X^-, \notag\\
d^{\text{c}}&=-\frac{s_{2\theta_\alpha} s_{2\theta_\delta}}{4}
\left[\frac{2\boldsymbol{k}^2}{\rho_0} \left(J_1^\beta+J^d_s-J_3^\beta-J_s\right) 
+ s_{\theta_\beta}^2(2\epsilon_X^- -\epsilon_{\text{cap}})\right]
-\frac{N}{2},\notag\\ 
e^{\text{c}}&=-\frac{L}{2}, \quad f^{\text{c}}=\frac{N}{2},    
\label{matrixelements2}
\end{align} 
with  
\begin{align} 
J_3^{\alpha}&=c_{\theta_\alpha}^2 J^d_s+s_{\theta_\alpha}^2 J_s, \quad
J_1^\beta=c_{\theta_\beta}^2 J_s+s_{\theta_\beta}^2 J^d_s, \quad  
J_2^\beta=\frac{s_{2\theta_\beta}}{2}  (J^d_s-J_s), \quad 
J_3^\beta=c_{\theta_\beta}^2 J^d_s+s_{\theta_\beta}^2 J_s,  
\nonumber\\ 
L&=-\frac{s_{2\theta_\beta}}{2} \left[s_{\theta_\delta} s_{2\theta_\alpha} (2\epsilon_X^- -\epsilon_{\text{cap}})
+c_{\theta_\alpha}\frac{\Delta_{\text{SAS}}}{\Delta_0}\epsilon_{\alpha}\right],\quad
\epsilon_{\alpha}=4c^2_{\theta_\alpha} \epsilon _X^-+2s_{\theta_\alpha}^2\epsilon_{\text{cap}},
\notag\\
N&=\frac{s_{2\theta_\delta} s_{2\theta_\alpha}s_{\theta_\beta}^2}{2} 
 (2\epsilon_X^- -\epsilon_{\text{cap}}) 
+\frac{\Delta_{\text{SAS}}}{\Delta_0}(c_{\theta_\delta}c_{\theta_\alpha}s^2_{\theta_\beta}  \epsilon_{\alpha}+\Delta_Z),
\label{matrixelements3}
\end{align}   
where we denote $s_{2\theta_{\alpha}}=\sin{2\theta_{\alpha}}$, $s_{2\theta_{\beta}}=\sin {2\theta_{\beta}}$, 
and $s_{2\theta_{\delta}}=\sin{2\theta_{\delta}}$.

It can be verified that the effective Hamiltonian \eqref{scantedhamiltonian1} and \eqref{scantedhamiltonian2}
reproduces the effective Hamiltonian in the spin phase \eqref{totalspinhamiltonian} 
by taking the limit  $\alpha,\beta\rightarrow0$.   
On the other hand, we reproduce the effective Hamiltonian in the pseudospin phase \eqref{totalppinhamiltonian}
by taking the limit $\alpha\rightarrow1$ in \eqref{scantedhamiltonian1} and \eqref{scantedhamiltonian2}.

The effective Hamiltonian in the CAF phase is too complicated to make a further analysis. 
We take the limit  $\Delta_{\text{SAS}}\rightarrow0$ to examine 
if some simplified formulas are obtained. 
In particular we would like to seek  gapless modes. 
Such gapless modes will play an important role in driving
the interlayer coherence in the CAF phase.  
In this limit, we have:
\begin{align}
\cos{\theta_\beta}&=\frac{\Delta_{\text{SAS}}}{\Delta_{\text{Z}}}, \quad 
\sin{\theta_\beta}=\pm \sqrt{1-\left(\frac{\Delta_{\text{SAS}}}{\Delta_{\text{Z}}}\right)^2}, \quad
\cos{\theta_\delta}=\cos{\theta_\alpha}, \quad \sin{\theta_\delta}=\sin{\theta_\alpha},\notag\\
\alpha^2&=|\sigma_0|.
\label{limitsincos}
\end{align}
  
From \eqref{orderparameter1} and \eqref{limitsincos}, the classical ground state reads:
\begin{align}
& \mathcal{S}_{z}^{0}=1-|\sigma_{0}|,\quad \mathcal{P}_{z}^{0}
=\sigma_{0},\quad \mathcal{R}_{xx}^{0}=\text{sgn}(\sigma_{0})\mathcal{R}_{yy}^{0},
\quad
 \mathcal{R}_{yy}^{0}=-\sqrt{|\sigma_{0}|(1-|\sigma_{0}|)},
\label{orderparameter}
\end{align}
all others being zero.
We assume $\sigma_0>0$  for definiteness. 
The transformation \eqref{su4rotation} has a simple expression:
\begin{align}
U^\dagger=
\left( 
\begin{array}{cccc}
1 & 0 & 0 & 0  \\
0 & \sqrt{1-|\sigma_0|} & \sqrt{|\sigma_0|} & 0    \\
0 & -\sqrt{|\sigma_0|} & \sqrt{1-|\sigma_0|} & 0  \\
0 & 0 & 0 & 1   \\
\end{array} 
\right),\label{cafunitarytransformation}
\end{align}
We find $\bar{\boldsymbol{n}}=U^\dagger{\boldsymbol{n}}$ is of the form 
$(n^{\text{f}\uparrow}, 
n^{\text{S}}_{\text{f}\downarrow\text{b}\uparrow}, 
n^{\text{A}}_{\text{f}\downarrow\text{b}\uparrow},  
n^{\text{b}\downarrow})^t$ by setting
\begin{align}    
n^{\text{S}}_{\text{f}\downarrow\text{b}\uparrow}=
(\sqrt{1-|\sigma_0|} n^{\text{f}\downarrow} +\sqrt{|\sigma_0|} n^{\text{b}\uparrow} ), \quad
n^{\text{A}}_{\text{f}\downarrow\text{b}\uparrow}=
(-\sqrt{|\sigma_0|} n^{\text{f}\downarrow}+\sqrt{1-|\sigma_0|} n^{\text{b}\uparrow} ).
\label{cafgroundstate}
\end{align}
Consequently, the ground state is such that $|n^{\text{f}\uparrow}\rangle$
 and $|n^{\text{A}}_{\text{f}\downarrow\text{b}\uparrow}\rangle$
 are filled up: The NG modes $\eta_1$ and $\eta_3$ describe
an excitation from the state $|n^{\text{f}\uparrow}\rangle$
 to $|n^{\text{S}}_{\text{f}\downarrow\text{b}\uparrow}\rangle$
and $|n^{\text{b}\downarrow}\rangle$, respectively, 
while the NG modes $\eta_2$ and $\eta_4$ describe
an excitation from the state $|n^{\text{A}}_{\text{f}\downarrow\text{b}\uparrow}\rangle$
to $|n^{\text{b}\downarrow}\rangle$ 
and $|n^{\text{S}}_{\text{f}\downarrow\text{b}\uparrow}\rangle$, respectively. 
A similar analysis can be done for $\sigma_0 < 0$: $|n^{\text{b}\uparrow}\rangle$ and
$|n^{\text{S}}_{\text{f}\uparrow\text{b}\downarrow}\rangle$  are filled up, where
\begin{align}    
n^{\text{S}}_{\text{f}\uparrow\text{b}\downarrow}=
(\sqrt{1-|\sigma_0|} n^{\text{f}\uparrow} +\sqrt{|\sigma_0|} n^{\text{b}\downarrow} ), \quad
n^{\text{A}}_{\text{f}\downarrow\text{b}\uparrow}=
(-\sqrt{|\sigma_0|} n^{\text{f}\uparrow}+\sqrt{1-|\sigma_0|} n^{\text{b}\downarrow} ),
\label{cafgroundstate2}
\end{align}
and the gapless mode $\eta_4$ describes an excitation from the state $|n^{\text{S}}_{\text{f}\uparrow\text{b}\downarrow}\rangle$ 
to $|n^{\text{A}}_{\text{f}\uparrow\text{b}\downarrow}\rangle$.

\begin{figure}[t]  
\begin{center} 
\centering
\includegraphics[width=70mm]{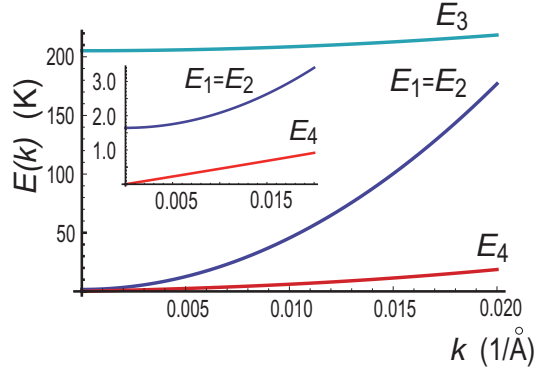}\hspace{1pc}%
\caption{\label{label}
 Dispersion relations \eqref{scmixingdispersions} for the four NG modes $E_i$.
The sample parameters are 
$d=231\AA$, $B\approx5.6\text{T}$, $\rho_0=2.7\times10^{15}\text{m}^{-2}$, and $\alpha=0.1$.
Inset: Dispersion relations near $k=0$. It is clear that $E_4(k)$ is linear.
}\label{newnu2ngmodedispersion}
\end{center}
\end{figure}

By using \eqref{limitsincos} with \eqref{scantedhamiltonian1}, and \eqref{scantedhamiltonian2}
with \eqref{matrixscaf}, \eqref{matrixelements1}, \eqref{matrixelements2}, and \eqref{matrixelements3},
we have the Hamiltonian
\begin{align} 
{H}=\sum_{i=1}^4 \int d^2 k 
E_{i}
\eta^{\text{c}\dagger}_{i,\boldsymbol{k}}\eta^{\text{c}}_{i,\boldsymbol{k}}, 
\label{limitscantedhamiltoniantotal}
\end{align}
together with the dispersion relations (Fig. \ref{newnu2ngmodedispersion}):
\begin{align}
E_1&=E_2=\frac{4\boldsymbol{k}^2}{\rho_0}J^\alpha_1+\Delta_{\text{Z}},  
\quad
E_{3}=\frac{4\boldsymbol{k}^2}{\rho_0}J^d_s +2\Delta_{\text{Z}}+8\cos^2\theta_{\alpha}\epsilon^-_X,\notag\\
E_{4}&=|\boldsymbol{k}| 
\sqrt{\frac{8J^d_s}{\rho_0}
\left(
\frac{2\boldsymbol{k}^2}{\rho_0}(\cos^2 2\theta_{\alpha}J^d_s+\sin^2 2\theta_{\alpha}J_s) +2\sin^2 2\theta_{\alpha}(\epsilon^-_D-\epsilon^-_X )
\right)},
\label{scmixingdispersions}
\end{align}
 where $\eta^{\text{c}}_{i,\boldsymbol{k}}$ ($i=1,2,3,4$) are the annihilation operators
\begin{align}
\eta^{\text{c}}_{1,\boldsymbol{k}}&=\frac{\check{\vartheta}_{1,\boldsymbol{k}}-
i\check{\sigma}_{1,\boldsymbol{k}}}{\sqrt{2}},\quad
\eta^{\text{c}}_{2,\boldsymbol{k}}=\frac{\check{\vartheta}_{2,\boldsymbol{k}}-
i\check{\sigma}_{2,\boldsymbol{k}}}{\sqrt{2}},\quad
\eta^{\text{c}}_{3,\boldsymbol{k}}=
\left(\frac{\rho_0}{4}\right)^{\frac{1}{2}}
\left(
\sigma_{3,\boldsymbol{k}}+i\vartheta_{3,\boldsymbol{k}}
\right),\notag\\
\eta^\text{c}_{4,\boldsymbol{k}}&=
\left(\frac{\rho_0}{4}\right)^{\frac{1}{2}}
\left(\left(\frac{\lambda^{\sigma_4}}{\lambda^{\vartheta_4}}\right)^{\frac{1}{4}}\sigma_{4,\boldsymbol{k}}
+i\left(\frac{\lambda^{\vartheta_4}}{\lambda^{\sigma_4}}\right)^{\frac{1}{4}}\vartheta_{4,\boldsymbol{k}}\right),
\label{cafannihilationmode} 
\end{align} 
with
\begin{align}
\lambda^{\vartheta_4}=
\frac{2\boldsymbol{k}^2}{\rho_0}J^d_s, \quad 
\lambda^{\sigma_4}=
\frac{2\boldsymbol{k}^2}{\rho_0}(\cos^2{2\theta_\alpha}J^d_s+\sin^2{2\theta_\alpha}J_s) 
+2\sin^2{2\theta_\alpha}(\epsilon^-_D-\epsilon^-_X ).  
\label{eigenvaluesc} 
\end{align}
The annihilation operators $\eta_{i,\boldsymbol{k}}$  satisfy the commutation relation, 
\begin{align}
\left[\eta^\text{c}_{i,\boldsymbol{k}},\eta^{\text{c}\dagger}_{j,\boldsymbol{k}^\prime}\right]
=\delta_{ij}\delta(\boldsymbol{k}-\boldsymbol{k}^\prime), 
\end{align}
with $i,j=1,2,3,4$.  

We summarize the NG modes in the CAF phase in the limit $\Delta_{\text{SAS}}\rightarrow0$. 
It is to be emphasized that there emerges one gapless mode, $\eta^\text{c}_{4,\boldsymbol{k}}$,   
reflecting the realization of the exact and its spontaneous breaking of 
the U(1) symmetry generated by $\frac{T_{yx}-T_{xy}}{\sqrt{2}}$.  
Furthermore, it has the linear dispersion relation as in (\ref{scmixingdispersions}), 
which leads to a superfluidity associated with this gapless mode. All other modes have gaps. 

\subsection{CAF phase in $\Delta_{\text{SAS}}\rightarrow0$ up to $\mathcal{O}(\Delta_{\text{SAS}}^3)$} 
\label{ngmodecafDSAS2}

We focus solely on the gapless mode $\eta^\text{c}_4$ ( or $\eta_4$ ) by neglecting all other gapped modes,
and derive the effective Hamiltonian for  $\eta_4$ up to $\mathcal{O}(\Delta _{\text{SAS}}^3)$. 
We assume $\sigma_0 > 0$ for  simplicity.  

The two $\text{CP}^3$ fields to be used in the perturbation theory are given by 
$\bar{\boldsymbol{n}}=U^\dagger{\boldsymbol{n}}$
 with \eqref{su4rotation} and \eqref{perturbativecp3parametrization}, or
\begin{equation}
\bar{\boldsymbol{n}}_{1}
=\left(\begin{array}{c}
1 \\ 
0 \\ 
0 \\ 
0
\end{array}\right) ,
\quad \bar{\boldsymbol{n}}_{2} 
=\left(\begin{array}{c}
0 \\ 
\eta_4 \\ 
1-\frac{1}{2}|\eta_4|^2  \\ 
0
\end{array}\right),  \label{perturbativeeta4parametrization}
\end{equation}

Using \eqref{ImbalParam}, we can exactly determine $\beta$ as  
\begin{equation}
\beta^2=\frac{\Delta _{\text{SAS}}^{2}\alpha ^{2}+\Delta _{\text{Z}}^{2}(1-\alpha ^{2})}
{\Delta _{\text{SAS}}^{2}\alpha ^{4}+\sigma_0^2\Delta _{\text{Z}}^{2}(1-\alpha ^{2})}\sigma_0^2.
\label{exactbeta2}
\end{equation}
Note that in the limit $\Delta _{\text{SAS}}\rightarrow0$ we obtain $\beta\rightarrow1$,
which is in accord with our previous calculations.
Substituting \eqref{exactbeta2} into \eqref{vareq1}, we find
\begin{equation}
\Delta _{\text{Z}}^{2}=
\frac{\Delta _{\text{SAS}}^{2}\alpha ^{4}+\sigma_0^2\Delta _{\text{Z}}^{2}(1-\alpha ^{2})}
{\alpha ^{2}(\alpha ^{2}-\sigma_0^2)}+4\epsilon^-_X
\frac{\sigma_0^2-\alpha^4}{\alpha^3}\frac{\sqrt{\Delta _{\text{SAS}}^{2}\alpha ^{2}+\Delta _{\text{Z}}^{2}(1-\alpha ^{2})}}{\sqrt{\alpha^2-\sigma_0^2}}.
\label{deltaz2}
\end{equation}
The relation \eqref{deltaz2} determines the value of $\alpha ^{2}$ as a function of 
$\Delta _{\text{Z}}$, $\Delta _{\text{SAS}}$, and $\sigma_0$.
 Substituting this value into \eqref{exactbeta2} we
obtain $\beta^{2}$ as a function of 
$\Delta _{\text{Z}}$, $\Delta _{\text{SAS}}$, $\sigma_0$. 
We have thus summarized our problem into a single equation \eqref{deltaz2}.
When $\Delta _{\text{SAS}}$ is exactly zero, 
\eqref{deltaz2} yields the relation $\alpha ^{2}=|\sigma_0|$. 
Therefore, for weak tunnelings, we search for a solution in the form
\begin{equation}
\alpha^2=|\sigma_0|+\lambda\Delta _{\text{SAS}}^2+\mathcal{O}(\Delta _{\text{SAS}}^4),
\label{alpha2solution} 
\end{equation}
where we expect $\lambda$ to be a constant.
In order to find the value of $\lambda$ we use \eqref{alpha2solution} 
 and expand the relevant combinations in powers of $\Delta _{\text{SAS}}^2$. 
In particular, for the first and  second terms of \eqref{deltaz2} we find
\begin{align}
&\frac{\Delta _{\text{SAS}}^{2}\alpha ^{4}+\sigma_0^2\Delta _{\text{Z}}^{2}(1-\alpha ^{2})}
{\alpha ^{2}(\alpha ^{2}-\sigma_0^2)}=\Delta _{\text{Z}}^{2}\left[
1+\frac{(1-\lambda\Delta _{\text{Z}}^{2} )\Delta _{\text{SAS}}^{2}}{(1-|\sigma_0|)\Delta _{\text{Z}}^{2}}
-\frac{\lambda (2-|\sigma_0| )}{|\sigma_0|(1-|\sigma_0|)}\Delta _{\text{SAS}}^{2}
\right]+\mathcal{O}(\Delta _{\text{SAS}}^4),\notag\\
&4\epsilon^-_X
\frac{\sigma_0^2-\alpha^4}{\alpha^3}\frac{\sqrt{\Delta _{\text{SAS}}^{2}\alpha ^{2}+\Delta _{\text{Z}}^{2}(1-\alpha ^{2})}}{\sqrt{\alpha^2-\sigma_0^2}}=
-\lambda \frac{8\epsilon^-_X \Delta _{\text{Z}}}{|\sigma_0|}\Delta _{\text{SAS}}^{2}+\mathcal{O}(\Delta _{\text{SAS}}^4).
\end{align}
Substituting these into \eqref{deltaz2} we obtain
\begin{align}
\Delta _{\text{Z}}^{2}=\Delta _{\text{Z}}^{2}\left[
1+\frac{(1-\lambda\Delta _{\text{Z}}^{2} )\Delta _{\text{SAS}}^{2}}{(1-|\sigma_0|)\Delta _{\text{Z}}^{2}}
-\frac{\lambda (2-|\sigma_0| )}{|\sigma_0|(1-|\sigma_0|)}\Delta _{\text{SAS}}^{2}
\right]
-\lambda \frac{8\epsilon^-_X \Delta _{\text{Z}}}{|\sigma_0|}\Delta _{\text{SAS}}^{2}+\mathcal{O}(\Delta _{\text{SAS}}^4).
\end{align}
The lowest terms $\Delta _{\text{SAS}}^{0}$ disappear automatically. Requiring the 
$\Delta _{\text{SAS}}^{2}$-terms to vanish, we obtain
\begin{equation}
\lambda=\frac{1}{\Delta _{\text{Z}}}\frac{|\sigma_0|}{2(\Delta _{\text{Z}}+4\epsilon^-_X(1-|\sigma_0|))},
\end{equation}
and for $\alpha^2$ we summarize as
\begin{equation}
\alpha^2=|\sigma_0|\left(
1+\frac{\Delta _{\text{Z}}}{2(\Delta _{\text{Z}}+4\epsilon^-_X(1-|\sigma_0|))}
\frac{\Delta _{\text{SAS}}^{2}}{\Delta _{\text{Z}}^{2}}
\right)+\mathcal{O}(\Delta _{\text{SAS}}^4).\label{weakalpha2}
\end{equation}
Using this in \eqref{exactbeta2} we come to
\begin{equation}
\beta^2=1-\frac{\Delta _{\text{SAS}}^{2}}{\Delta _{\text{Z}}^{2}}
+\mathcal{O}(\Delta _{\text{SAS}}^4).\label{weakbeta2}
\end{equation}
Finally, using \eqref{weakalpha2} and \eqref{weakbeta2} in
\eqref{sincosdefinition} and  \eqref{vareq2}, we find:
\begin{align}
\sin^2\theta_\delta&=|\sigma_0|\left(
1+\frac{\Delta _{\text{Z}}+8\epsilon^-_X(1-|\sigma_0|))}{2(\Delta _{\text{Z}}+4\epsilon^-_X(1-|\sigma_0|))}
\frac{\Delta _{\text{SAS}}^{2}}{\Delta _{\text{Z}}^{2}}
\right)+\mathcal{O}(\Delta _{\text{SAS}}^4)\label{weaksindelta}\\
\Delta _{\text{bias}}&=\text{sgn}(\sigma_0)\Delta _{\text{Z}}
\left[
1+\frac{4\epsilon^-_X+8(\epsilon^-_D-\epsilon^-_X)|\sigma_0|}{\Delta _{\text{Z}}}
-\frac{1}{2}\frac{\Delta _{\text{SAS}}^{2}}{\Delta _{\text{Z}}^{2}}
\right]+\mathcal{O}(\Delta _{\text{SAS}}^4),\label{weakbias}
\end{align}
respectively. 
Then by using \eqref{weakalpha2}, \eqref{weakbeta2}, \eqref{weaksindelta}, and \eqref{weakbias} with \eqref{su4effectivehamiltonian1},
we obtain the effective Hamiltonian for the gapless mode $\eta_4$ ($\sigma_4$ and $\vartheta_4$): 
\begin{equation}
\mathcal{H}=\frac{J_{\vartheta_4}}{2}(\nabla\vartheta_4)^2+\frac{J_{\sigma_4}}{2}(\nabla\sigma_4)^2+
4\rho_0(\epsilon^-_D-\epsilon^-_X)|\sigma_0|\left(1-|\sigma_0|-\frac{1}{2}\frac{\Delta _{\text{SAS}}^{2}}{\Delta _{\text{Z}}^{2}}\right),  
\end{equation}
with 
\begin{equation} 
J_{\vartheta_4}=2\left(J^d_s+J^-_s\frac{\Delta _{\text{SAS}}^{2}}{\Delta _{\text{Z}}^{2}}\right), \quad
J_{\sigma_4}=2\left(J^d_s+8J^-_s|\sigma_0|(1-|\sigma_0|)
+J^-_s(1-4|\sigma_0|)\frac{\Delta _{\text{SAS}}^{2}}{\Delta _{\text{Z}}^{2}}\right).
\end{equation}
Taking $\Delta _{\text{SAS}}^{2}=0$,  we reproduce the previously
calculated expressions \eqref{limitscantedhamiltoniantotal} and 
\eqref{scmixingdispersions}.

We wish to derive the effective Hamiltonian for the nonperturbative analysis of the phase
field $\vartheta (\boldsymbol{x})$. For this purpose, it is necessary to start with the parameterization of the Grassmannian field valid for
arbitrary values of $\vartheta (\boldsymbol{x}).$ We make an ansatz
\begin{equation}
\boldsymbol{n}_{2}
=\left(\begin{array}{c}
0 \\ 
-e^{+i\vartheta (\boldsymbol{x})}\sqrt{\sigma (\boldsymbol{x})} \\ 
\sqrt{1-\sigma (\boldsymbol{x})} \\ 
0
\end{array}\right)=
e^{i\sigma_0\vartheta (\boldsymbol{x})}\left(\begin{array}{c}
0 \\ 
-e^{+i(1-\sigma_0)\vartheta (\boldsymbol{x})}\sqrt{\sigma (\boldsymbol{x})} \\ 
e^{-i\sigma_0\vartheta (\boldsymbol{x})}\sqrt{1-\sigma (\boldsymbol{x})} \\ 
0
\end{array}\right). \label{nonperturbativeparametrization2} 
\end{equation}
We expand it around  $\vartheta (\boldsymbol{x})=0$ and $\sigma (\boldsymbol{x})=\sigma_0$
 by setting $\delta\sigma (\boldsymbol{x})\equiv\sigma (\boldsymbol{x})-\sigma_0$. 
Up to the linear orders in  $\vartheta (\boldsymbol{x})$ and $\delta\sigma (\boldsymbol{x})$,
it is straightforward to show that
\begin{align}
e^{+i(1-\sigma_0)\vartheta (\boldsymbol{x})}\sqrt{\sigma (\boldsymbol{x})}&=
\sqrt{\sigma_0}-\sqrt{1-\sigma_0}\eta_4(\boldsymbol{x}),\notag\\
e^{-i\sigma_0\vartheta (\boldsymbol{x})}\sqrt{1-\sigma (\boldsymbol{x})}&=
\sqrt{1-\sigma_0}+\sqrt{\sigma_0}\eta_4(\boldsymbol{x}),
\end{align}
where we have set
\begin{equation}
\eta_4(\boldsymbol{x})=-\frac{\sigma (\boldsymbol{x})-\sigma_0}{2\sqrt{\sigma_0(1-\sigma_0)}}
-i\vartheta (\boldsymbol{x})\sqrt{\sigma_0(1-\sigma_0)}.
\end{equation}
By requiring the commutation relation \eqref{nu2etaccr}, we find
\begin{equation} 
\frac{\rho_0}{2}\left[
\sigma (\boldsymbol{x}),\vartheta (\boldsymbol{y})
\right]=i\delta(\boldsymbol{x}-\boldsymbol{y})
\end{equation}
We have shown that the $\text{CP}^3$ field \eqref{nonperturbativeparametrization2} 
 is reduced to $\boldsymbol{n}_{2}$ 
in \eqref{perturbativeeta4parametrization} in the linear order of the perturbation fields,
apart from the U(1) factor $e^{-i\sigma_0\vartheta (\boldsymbol{x})}$. We may drop it off the parameterization since the $\text{CP}^3$ field is 
defined up to such a U(1) factor. Indeed, such a factor does not contribute to the isospin fields.

Here we parameterize the $\text{CP}^{3}$ fields as 
\begin{equation}
\boldsymbol{n}_{1}
=\left(\begin{array}{c}
1 \\ 
0 \\ 
0 \\ 
0
\end{array}\right) ,
\quad \boldsymbol{n}_{2}
=\left(\begin{array}{c}
0 \\ 
-e^{+i\vartheta (\boldsymbol{x})/2}\sqrt{\sigma (\boldsymbol{x})} \\ 
e^{-i\vartheta (\boldsymbol{x})/2}\sqrt{1-\sigma (\boldsymbol{x})} \\ 
0
\end{array}\right) ,  \label{positivecp3}
\end{equation}
for $\sigma (\boldsymbol{x})>0$, and 
\begin{equation}
\boldsymbol{n}_{1}=
\left(\begin{array}{c}
0 \\ 
0 \\ 
1 \\ 
0
\end{array}\right) ,\quad \boldsymbol{n}_{2}=
\left(\begin{array}{c}
e^{+i\vartheta (\boldsymbol{x})/2}\sqrt{1+\sigma (\boldsymbol{x})} \\ 
0 \\ 
0 \\ 
e^{-i\vartheta (\boldsymbol{x})/2}\sqrt{-\sigma (\boldsymbol{x})}
\end{array}\right) .  \label{negativecp3}
\end{equation}
for $\sigma (\boldsymbol{x})<0$. The isospin density fields are expressed in
terms of $\sigma (\boldsymbol{x})$ and $\vartheta (\boldsymbol{x})$: 
\begin{align}
& \mathcal{S}_{z}(\boldsymbol{x})=1-|\sigma (\boldsymbol{x})|,
\quad \mathcal{P}_{z}(\boldsymbol{x})=\sigma (\boldsymbol{x}),  \notag \\
& \mathcal{R}_{yy}(\boldsymbol{x})
=\text{sgn}(\sigma_{0})\mathcal{R}_{xx}(\boldsymbol{x})
=-\sqrt{|\sigma (\boldsymbol{x})|(1-|\sigma (\boldsymbol{x})|)}\cos \vartheta (\boldsymbol{x}),  \notag \\
& \mathcal{R}_{yx}(\boldsymbol{x})=-\text{sgn}(\sigma_{0})\mathcal{R}_{xy}(\boldsymbol{x})
=-\sqrt{|\sigma (\boldsymbol{x})|(1-|\sigma (\boldsymbol{x})|)}\sin \vartheta (\boldsymbol{x}),  \label{Isospin}
\end{align}
with all others being zero. The ground-state expectation values are 
$\langle\sigma (\boldsymbol{x})\rangle =\sigma_{0}$, 
$\langle \vartheta (\boldsymbol{x})\rangle =0$, with which the order parameters 
\eqref{orderparameter} are reproduced from (\ref{Isospin}). It is notable
that the fluctuations of the phase field $\vartheta (\boldsymbol{x})$ affect
both the spin and pseudospin components of the $R$-spin. This is very different
from the spin wave in the monolayer QH system or the pseudospin wave in the
bilayer QH system at $\nu =1$. Hence we call it the entangled
spin-pseudospin phase field $\vartheta (\boldsymbol{x})$.

By substituting (\ref{Isospin}) into \eqref{su4effectivehamiltonian1}, 
apart from irrelevant
constant terms, the resulting effective Hamiltonian is: 
\begin{equation}
\mathcal{H}_{\text{eff}}=\frac{J_{\vartheta}}{2}\left( \nabla \vartheta
\right)^{2}+\frac{J_{\sigma}}{2}\left( \nabla {\sigma}\right)^{2}
+\rho_{\Phi}\epsilon_{\text{cap}}^{\nu =1}(\sigma -\sigma_{0})^{2},
\label{EffecHamil}
\end{equation}
where we have used
\begin{align}
\Delta_{\text{bias}}&=\text{sgn}(\sigma_{0})\left[
\Delta_{\text{Z}}+4\epsilon_{X}^{-}+2\epsilon_{\text{cap}}^{\nu =1}|\sigma_0|
\right],\\
J_{\sigma}&=4J_{s}+\frac{(2|\sigma_{0}|-1)^{2}}{|\sigma_{0}|(1-|\sigma_{0}|)}J_{s}^{d},\quad
J_{\vartheta}=4{J_{s}^{d}}|\sigma_{0}|(1-|\sigma_{0}|).
\end{align}
When we require the equal-time commutation relation, 
\begin{equation}
\frac{\rho_{0}}{2}\left[ \sigma (\boldsymbol{x}),\vartheta (\boldsymbol{y})\right] 
=i\delta (\boldsymbol{x}-\boldsymbol{y}),  \label{CR}
\end{equation}
the Hamiltonian (\ref{EffecHamil}) is second quantized, and it has the
linear dispersion relation 
\begin{equation}
E_{\boldsymbol{k}}=|\boldsymbol{k}|\sqrt{\frac{2J_{\vartheta}}{\rho_{0}}
\left( \frac{2J_{\sigma}}{\rho_{0}}\boldsymbol{k}^{2}
+2\epsilon_{\text{cap}}^{\nu =1}\right)}.
\end{equation}
This agrees with $E_4$ in Eq. \eqref{scmixingdispersions}. 
It should be emphasized that 
the effective Hamiltonian (\ref{EffecHamil}) is valid in all orders of the
phase field $\vartheta (\boldsymbol{x})$. It may be regarded as a classical
Hamiltonian as well, where (\ref{CR}) should be replaced with the
corresponding Poisson bracket.

The effective Hamiltonian (\ref{EffecHamil}) for $\vartheta (\boldsymbol{x})$
and $\sigma (\boldsymbol{x})$ reminds us of the one that governs the
Josephson effect at $\nu =1$. The main difference is the absence of the
tunneling term, which implies that there exists no Josephson tunneling. We have
shown that the effective Hamiltonian is correct up to 
$\mathcal{O}(\Delta_{\text{SAS}}^{3})$ as $\Delta_{\text{SAS}}\rightarrow 0$. 
Nevertheless, the Josephson supercurrent is present within the layer, which is our main issue.

By using the Hamiltonian (\ref{EffecHamil}) and the commutation relation \eqref{CR}, 
we obtain the equations of motion: 
\begin{align}
\hbar \partial_{t}\vartheta (\boldsymbol{x})& 
=\frac{2J_{\sigma}}{\rho_{0}}\nabla^{2}\sigma (\boldsymbol{x})
-2\epsilon_{\text{cap}}^{\nu =1}(\sigma (\boldsymbol{x})
-\sigma_{0}),  \label{heisenbergeom1} \\
\hbar \partial_{t}\sigma (\boldsymbol{x})& =-\frac{2J_{\vartheta}}
{\rho_{0}}\nabla^{2}\vartheta (\boldsymbol{x}).  \label{heisenbergeom2}
\end{align}
\subsection{Josephson supercurrents in the CAF phase}
\label{josephsonsupercurrentsnu2}
We now study the electric Josephson supercurrent carried by the gapless mode $\vartheta (\boldsymbol{x})$ in the CAF phase, 
where the further analysis goes in parallel with that given for $\nu=1.$ 
 
The electron densities are 
$\rho_{e}^{\text{f}(\text{b})}=-{e\rho_{0}}\left( 1\pm \mathcal{P}_{z}\right) /2
=-{e\rho_{0}}\left( 1\pm \sigma (\boldsymbol{x})\right) /2$ on each layer. 
Taking the time derivative and using \eqref{heisenbergeom2}, we find 
\begin{equation}
\partial_{t}\rho_{e}^{\text{f}}=-\partial_{t}\rho_{e}^{\text{b}} 
=\frac{eJ_{\vartheta}}{\hbar}\nabla^{2}\vartheta (\boldsymbol{x}).
\label{continuityequation2} 
\end{equation}
The time derivative of the charge is associated with the current via the
continuity equation, $\partial_{t}\rho_{e}^{\text{f}(\text{b})}
=\partial_{i}\mathcal{J}_{i}^{\text{f}(\text{b})}$. We thus identify 
$\mathcal{J}_{i}^{\text{f(b)}}=\pm \mathcal{J}_{i}^{\text{Jos}}(\boldsymbol{x})+$constant, 
where
\begin{equation}
\mathcal{J}_{i}^{\text{Jos}}(\boldsymbol{x})\equiv \frac{eJ_{\vartheta}}{\hbar}
\partial_{i}\vartheta (\boldsymbol{x}).  \label{phasecurrent2}  
\end{equation}
Consequently, the current $\mathcal{J}_{x}^{\text{Jos}}(\boldsymbol{x})$
flows when there exists inhomogeneity in the phase $\vartheta (\boldsymbol{x})$. 
Such a current is precisely the Josephson supercurrent. It is intriguing that the current does not flow in the balanced
system since $J_{\vartheta}=0$ at $\sigma_{0}=0$.
\subsection{Quantum Hall effects in the CAF phase} 
\label{qhenu2}
\begin{figure}[t]
\begin{center}
\centering
\includegraphics[width=0.85\textwidth]{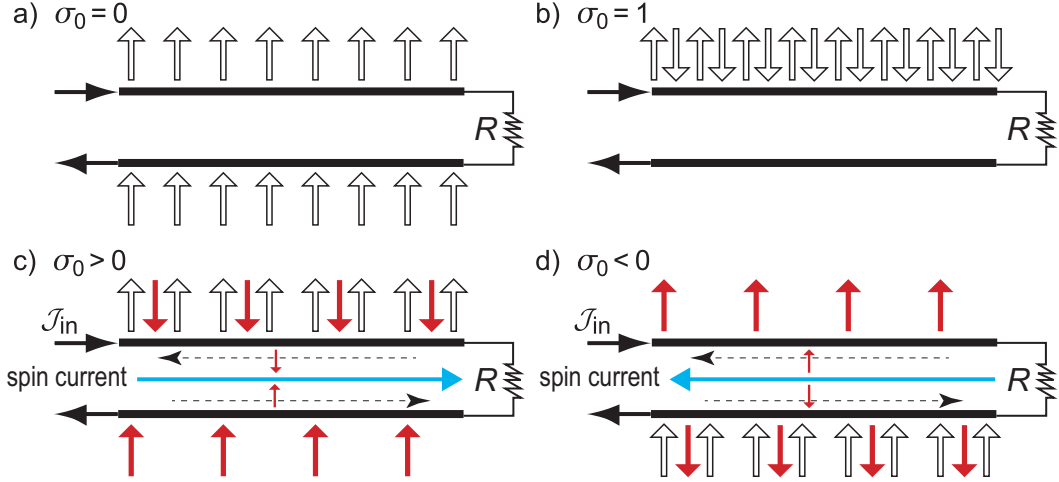}
\end{center}
\caption{  Schematic illustration of the spin supercurrent
flowing along the $x$-axis in the counterflow geometry. (a) All spins are
polarized into the positive $z$ axis due to the Zeeman effect at $\protect\sigma_{0}=0$. 
No spin current flows. (b) All electrons belong to the front
layer at $\protect\sigma_{0}=1$. No spin current flows. (c) In the CAF
phase for $\protect\sigma_{0}>0$, some up-spin electrons are moved from the
back layer to the front layer by flipping spins. An NG mode appears
associated with this charge-spin transfer. The interlayer phase
difference $\protect\vartheta (\boldsymbol{x})$ is created by feeding a 
charge current $\mathcal{J}_{\text{in}}$ to the front layer, which also
drives the spin current. Electrons flow in each layer as indicated by the
dotted horizontal arrows, and the spin current flows as indicated by the
solid horizontal arrow. (d) In the CAF phase for $\protect\sigma_{0}<0$,
similar phenomena occur but the direction of the spin current becomes
opposite.}
\label{nu2spincurrentfigure} 
\end{figure}

Let us inject the current $\mathcal{J}_{\text{in}}$ into the $x$ direction
of the bilayer sample, and assume the system to be homogeneous in the $y$
direction (Fig. \ref{nu2spincurrentfigure}). By applying the same argument as given in
Sect. \ref{qhenu1}, we show the anomalous Hall resistance behaviours affected by the phase coherence in the CAF phase.

The current for each layer is the sum of the Hall current and the Josephson current, 
\begin{equation}
\mathcal{J}_{x}^{\text{f}}(x)=\frac{\nu}{R_{\text{K}}}
\frac{\rho_{0}^{\text{f}}}{\rho_{0}}E_{y}+\mathcal{J}_{x}^{\text{Jos}},
\quad \mathcal{J}_{x}^{\text{b}}(x)=\frac{\nu}{R_{\text{K}}}\frac{\rho_{0}^{\text{b}}}
{\rho_{0}}E_{y}-\mathcal{J}_{x}^{\text{Jos}}.  \label{totalcurrent2}
\end{equation}
We apply these formulas to analyze the counterflow and drag experiments without tunneling. 
With the same argument as given in Sect. \ref{qhenu1}, we have 
\begin{equation}
R_{xy}^{\text{f}}\equiv \frac{E_{y}^{\text{f}}}{\mathcal{J}_{x}^{\text{f}}}=0,
\qquad 
R_{xy}^{\text{b}}\equiv \frac{E_{y}^{\text{b}}}{\mathcal{J}_{x}^{\text{b}}}=0  \label{counterflowanomalous2}
\end{equation}
in the counterflow experiment. 
All the input current is carried by the Josephson supercurrent, 
$\mathcal{J}_{x}^{\text{Jos}}=\mathcal{J}_{\text{in}}$. It generates such an
inhomogeneous phase field that 
$\vartheta (\boldsymbol{x})=(\hbar/eJ_{\vartheta})\mathcal{J}_{\text{in}}x$.

On the other hand, in the drag experiment, we have  
$\mathcal{J}_{\text{in}}=\mathcal{J}_{x}^{\text{f}}=(\nu /R_{\text{K}})E_{y}$, or 
\begin{equation}
R_{xy}^{\text{f}}\equiv {\frac{E_{y}^{\text{f}}}{\mathcal{J}_{x}^{\text{f}}}=
\frac{R_{\text{K}}}{\nu}=}\frac{1}{2}R_{\text{K}}\qquad \text{at\quad}\nu=2.
 \label{draganomalous2} 
\end{equation}  
A part of the input current is carried by the Josephson supercurrent, 
$\mathcal{J}_{x}^{\text{Jos}}=\frac{1}{2}(1-\sigma_{0})\mathcal{J}_{\text{in}}$.

In conclusion, we predict the
anomalous Hall resistance (\ref{counterflowanomalous2}) and (\ref{draganomalous2}) 
in the CAF phase at $\nu =2$ by carrying out similar
experiments\cite{Kellog1,Tutuc,Kellog2} due to Kellogg {et al}. and
Tutuc {et al}. in the imbalanced configuration ($\sigma_{0}\neq 0$).
\subsection{Spin Josephson supercurrent in the CAF phase} 
\label{spinjosephsonsupercurrentsnu2}
An intriguing feature of the CAF phase is that the phase field $\vartheta (\boldsymbol{x})$ 
describes the entangled spin-pseudospin coherence according
to the basic formula (\ref{Isospin}).  

Up to $\mathcal{O}((\sigma -\sigma_{0})^{2})$, we have 
$\mathcal{S}_{z}=1-|\sigma (\boldsymbol{x})|$, and we obtain 
\begin{align}
\partial_{t}\rho_{\text{b}\uparrow}^{\text{spin}}& =\partial_{t}
\rho_{\text{f}\downarrow}^{\text{spin}}=\frac{J_{\vartheta}}{4}[1+\text{sgn}
(\sigma_{0})]\partial_{x}^{2}\vartheta (\boldsymbol{x}), \\
\partial_{t}\rho_{\text{f}\uparrow}^{\text{spin}}& =\partial_{t}
\rho_{\text{b}\downarrow}^{\text{spin}}=-\frac{J_{\vartheta}}{4}[1-\text{sgn}(\sigma_{0})]\partial_{x}^{2}\vartheta (\boldsymbol{x}).
\label{rhotderivative}
\end{align}
The time derivative of the spin is associated with the spin current via the
continuity equation, 
$\partial_{t}\rho_{\alpha}^{\text{spin}}(\boldsymbol{x})
=\partial_{x}\mathcal{J}_{\alpha}^{\text{spin}}(\boldsymbol{x})$ for
each $\alpha $. We thus identify 
\begin{align}
\mathcal{J}_{\text{b}\uparrow}^{\text{spin}}(\boldsymbol{x})& 
=\mathcal{J}_{\text{f}\downarrow}^{\text{spin}}(\boldsymbol{x})
=\frac{J_{\vartheta}}{2}\partial_{x}\vartheta (\boldsymbol{x}),\quad \text{for}\ \sigma_{0}>0, \\
\mathcal{J}_{\text{f}\uparrow}^{\text{spin}}(\boldsymbol{x})& 
=\mathcal{J}_{\text{b}\downarrow}^{\text{spin}}(\boldsymbol{x})=-\frac{J_{\vartheta}}{2}
\partial_{x}\vartheta (\boldsymbol{x}),\quad \text{for}\ \sigma_{0}<0.
\end{align}
The spin current $\mathcal{J}_{\alpha}^{\text{spin}}(\boldsymbol{x})$ flows
along the $x$ axis, when there exists an inhomogeneous phase difference $\vartheta (\boldsymbol{x})$.

In the counterflow experiment, the total charge current along the $x$ axis is zero: 
$\mathcal{J}_{x}^{\text{f}}(\boldsymbol{x})+\mathcal{J}_{x}^{\text{b}}
(\boldsymbol{x})=0$. 
Consequently, the input current generates a pure spin current along the $x$-axis, 
\begin{equation}
\mathcal{J}_{x}^{\text{spin}}=\mathcal{J}_{\text{f}\uparrow}^{\text{spin}}+
\mathcal{J}_{\text{f}\downarrow}^{\text{spin}}
+\mathcal{J}_{\text{b}\uparrow}^{\text{spin}}+\mathcal{J}_{\text{b}\downarrow}^{\text{spin}}
=\text{sgn}(\sigma_{0})\frac{\hbar}{e}\mathcal{J}_{\text{in}}.
\end{equation}
This current is dissipationless since the dispersion relation is linear. It
is appropriate to call it a spin Josephson supercurrent. It is intriguing
that the spin current flows in the opposite directions for $\sigma_{0}>0$
and $\sigma_{0}<0$, as illustrated in Fig.\ref{nu2spincurrentfigure}. 
A comment is in order: The spin current only flows within the sample, since
spins are scattered in the resistor $R$ and spin directions become random
outside the sample.


\section{Conclusion}
In this paper, we have derived the effective Hamiltonian for the NG modes based on the Grassmannian formalism.
We have first reproduced the perturbative results on the dispersions and  coherence lengths obtained in Ref.\cite{yhama}.
We have then presented the effective theory describing 
the interlayer coherence in the bilayer QH system at $\nu=1,2$. 
The Grassmannian formalism shows a clear physical picture of the spontaneous development of an interlayer phase coherence. 
It is to be emphasized that the Grassmannian formalism enables us to analyze 
nonperturbative phase coherent phenomena such as the Josephson supercurrent.
The nonperturbative analysis was beyond the scope of Ref.\cite{yhama}.   
It has been argued\cite{Ezawa:2008ae} that the interlayer coherence is due to the Bose-Einstein condensation of composite bosons, which are single electrons bound to magnetic flux quanta.  
The composite bosons are described by the CP fields, from which the Grassmannian field is composed.

We have explored the phase-coherent phenomena in the bilayer system. 
At $\nu=1$, the interlayer phase coherence due to the pseudospin, governed by the  NG mode describing a pseudospin wave, is 
developed spontaneously.
On the other hand, the phase coherence in the CAF phase 
is the entangled spin-pseudospin phase coherence  governed by the NG mode $\vartheta (\boldsymbol{x})$
describing the $R$-spin according to the formula (\ref{Isospin}). 
We have predicted the anomalous Hall resistivity in the counterflow and drag experiments. 
It has been shown to exhibit precisely the same behaviour for $\nu =1$ and $\nu =2$.
The difference between them is that the supercurrent
flows both in  balanced and imbalanced systems at
$\nu =1$ but only in imbalanced systems at $\nu =2$. 
Furthermore,  a spin Josephson supercurrent flows in the CAF phase in the counterflow geometry,
but not for $\nu =1$.
In other words, the net spin current is nonzero for the CAF phase, 
while it is zero for $\nu=1$. This is due to the spin structure such that 
the spins are canted coherently and making antiferromagnetic correlations
between the two layers at $\nu=2$, while the spin is actually frozen and therefore all of the spins are
pointing to the positive $z$ axis in both layers at $\nu=1$ in the limit $\Delta_{\text{SAS}}\rightarrow0$.

\section*{Acknowledgment}
Y. Hama thanks Takahiro Morimoto and Akira Furusaki for useful discussions and comments.
This research was supported in part by JSPS Research Fellowships for Young
Scientists, and a Grant-in-Aid for Scientific Research from the Ministry of
Education, Culture, Sports, Science and Technology (MEXT) of Japan (No.
21540254).


%

\vfill\pagebreak

\appendix

\section{Appendix A SU(4) algebra}
\label{appendix}
The special unitary group SU(N) has $(N^2-1)$ generators. According to the standard notation from 
elementary particle physics\cite{Gell-Mann1}, we denote them as $\lambda_A$, $A=1,2,\ldots,N^2-1$,
which are represented by Hermitian, traceless, $N\times N$ matrices,  
and normalize them as
\begin{equation}
\text{Tr}(\lambda_A\lambda_B)=2\delta_{AB}.    
\label{orthogonality1}
\end{equation}
They are characterized by 
\begin{align} 
\left[ \lambda_A,\lambda_B\right]=2if_{ABC}\lambda_C, \quad
\{ \lambda_A,\lambda_B\}=\frac{4}{N}2d_{ABC}\lambda_C,
\end{align}
where $f_{ABC}$ and $d_{ABC}$ are the structure constants of SU(N). 
We have $\lambda_A=\tau_A$ (the Pauli matrix) with $f_{ABC}=\epsilon _{ABC}$ and $d_{ABC}=0$ in the case of SU(2). 
 
This standard representation is not convenient for our purpose because the spin group is 
$\text{SU}(2)\times \text{SU}(2)$ in the bilayer electron system with the four-component electron field as 
$\Psi=(\psi^{\text{f}\uparrow},\psi^{\text{f}\downarrow},\psi^{\text{b}\uparrow},\psi^{\text{b}\downarrow})$. 
Embedding $\text{SU}(2)\times \text{SU}(2)$ into SU(4) we define the spin matrix by 
\begin{equation}
\tau _{a}^{\text{spin}}=\left( 
\begin{array}{cc}
\tau _{a} & 0 \\ 
0 & \tau _{a}
\end{array}
\right) ,  \label{su4base1} 
\end{equation}
where $a=x,y,z$, and the pseudospin matrices by, 
\begin{align}
\tau _{x}^{\text{ppin}} =  \left( 
\begin{array}{cc}
0 & \boldsymbol{1}_{2} \\ 
\boldsymbol{1}_{2} & 0
\end{array}
\right), \quad 
\tau _{y}^{\text{ppin}}=\left( 
\begin{array}{cc}
0 & -i\boldsymbol{1}_{2} \\ 
i\boldsymbol{1}_{2} & 0
\end{array}
\right), \quad 
\tau _{z}^{\text{ppin}} =  \left( 
\begin{array}{cc}
\boldsymbol{1}_{2} & 0 \\ 
0 & -\boldsymbol{1}_{2}
\end{array}
\right) ,  \label{su4base2}
\end{align}
where $\boldsymbol{1}_{2}$ is the unit matrix in two dimensions. Nine remaining
matrices are simple products of the spin and pseudospin matrices: 
\begin{align}
\tau _{a}^{\text{spin}}\tau _{x}^{\text{ppin}} =\left( 
\begin{array}{cc}
0 & \tau _{a} \\ 
\tau _{a} & 0
\end{array}
\right), \quad 
 \tau _{a}^{\text{spin}}\tau _{y}^{\text{ppin}}=\left( 
\begin{array}{cc}
0 & -i\tau _{a} \\ 
i\tau _{a} & 0
\end{array}
\right), \quad
\tau _{a}^{\text{spin}}\tau _{z}^{\text{ppin}} =\left( 
\begin{array}{cc}
\tau _{a} & 0 \\ 
0 & -\tau _{a}
\end{array}
\right).  \label{su4base3}
\end{align}
We denote them $T_{a0}\equiv \frac{1}{2}\tau_a^{\text{spin}}$, $T_{0a}\equiv 
\frac{1}{2}\tau_a^{\text{ppin}}$,
$T_{ab}\equiv \frac{1}{2}\tau_a^{\text{spin}}\tau_b^{\text{ppin}}$. They satisfy the normalization condition 
\begin{equation}
\text{Tr}(T_{\mu \nu}T_{\gamma\delta})=\delta_{\mu\gamma}\delta_{\nu\delta},
\label{orthogonality2}
\end{equation}
and the commutation relations  
\begin{equation}
[T_{\mu\nu},T_{\gamma\delta}]=if_{\mu\nu,\gamma\delta,\mu^\prime\nu^\prime}T_{\mu^\prime\nu^\prime},
 \label{su4commutator1}
\end{equation}
where $f_{\mu\nu,\gamma\delta,\mu\prime\nu^\prime}$ is  
the SU(4) structure constant in the basis \eqref{su4base1}-\eqref{su4base3}.    
Greek indices run over $0,x,y,z$. 
 
From \eqref{su4rotation}, \eqref{sincosdefinition}, 
and  \eqref{perturbativecp3parametrization}, the explicit form of the isospin densities
in terms of $\eta_i$ is given by:
\begin{align}
\mathcal{I}_{0x}&=-\cos{\theta_\alpha}\sin{\theta_\beta}\mathcal{I}^{\text{c}}_{0x}+\cos{\theta_\alpha}\cos{\theta_\beta}\cos{\theta_\delta}\mathcal{I}^{\text{c}}_{0z}
-\sin{\theta_\alpha}\cos{\theta_\beta}\cos{\theta_\delta}\mathcal{I}^{\text{c}}_{xx}-\sin{\theta_\alpha}\sin{\theta_\beta}\mathcal{I}^{\text{c}}_{xz}
\notag\\
&-\cos{\theta_\alpha}\cos{\theta_\beta}\sin{\theta_\delta}\mathcal{I}^{\text{c}}_{yy}+\sin{\theta_\alpha}\cos{\theta_\beta}\sin{\theta_\delta}\mathcal{I}^{\text{c}}_{z0},\notag\\
\mathcal{I}_{0y}&=\cos{\theta_\delta}\mathcal{I}^{\text{c}}_{0y}+\sin{\theta_\delta}\mathcal{I}^{\text{c}}_{yz},\notag\\
\mathcal{I}_{0z}&=-\cos{\theta_\alpha}\cos{\theta_\beta}\mathcal{I}^{\text{c}}_{0x}-\cos{\theta_\alpha}\sin{\theta_\beta}\cos{\theta_\delta}\mathcal{I}^{\text{c}}_{0z}
+\sin{\theta_\alpha}\sin{\theta_\beta}\cos{\theta_\delta}\mathcal{I}^{\text{c}}_{xx}-\sin{\theta_\alpha}\cos{\theta_\beta}\mathcal{I}^{\text{c}}_{xz}
\notag\\
&+\cos{\theta_\alpha}\sin{\theta_\beta}\sin{\theta_\delta}\mathcal{I}^{\text{c}}_{yy}-\sin{\theta_\alpha}\sin{\theta_\beta}\sin{\theta_\delta}\mathcal{I}^{\text{c}}_{z0},\notag\\
\mathcal{I}_{x0}&=\cos{\theta_\delta}\mathcal{I}^{\text{c}}_{x0}-\sin{\theta_\delta}\mathcal{I}^{\text{c}}_{zx},\notag\\
\mathcal{I}_{xx}&=-\sin{\theta_\alpha}\cos{\theta_\beta}\mathcal{I}^{\text{c}}_{0x}-\sin{\theta_\alpha}\sin{\theta_\beta}\cos{\theta_\delta}\mathcal{I}^{\text{c}}_{0z}
-\cos{\theta_\alpha}\sin{\theta_\beta}\cos{\theta_\delta}\mathcal{I}^{\text{c}}_{xx}+\cos{\theta_\alpha}\cos{\theta_\beta}\mathcal{I}^{\text{c}}_{xz}
\notag\\
&+\sin{\theta_\alpha}\sin{\theta_\beta}\sin{\theta_\delta}\mathcal{I}^{\text{c}}_{yy}+\cos{\theta_\alpha}\sin{\theta_\beta}\sin{\theta_\delta}\mathcal{I}^{\text{c}}_{z0},\notag\\
\mathcal{I}_{xy}&=\mathcal{I}^{\text{c}}_{xy},\notag\\
\mathcal{I}_{xz}&=\sin{\theta_\alpha}\sin{\theta_\beta}\mathcal{I}^{\text{c}}_{0x}-\sin{\theta_\alpha}\cos{\theta_\beta}\cos{\theta_\delta}\mathcal{I}^{\text{c}}_{0z}
-\cos{\theta_\alpha}\cos{\theta_\beta}\cos{\theta_\delta}\mathcal{I}^{\text{c}}_{xx}
-\cos{\theta_\alpha}\sin{\theta_\beta}\mathcal{I}^{\text{c}}_{xz}
\notag\\
&+\sin{\theta_\alpha}\cos{\theta_\beta}\sin{\theta_\delta}\mathcal{I}^{\text{c}}_{yy}
+\cos{\theta_\alpha}\cos{\theta_\beta}\sin{\theta_\delta}\mathcal{I}^{\text{c}}_{z0},\notag\\
\mathcal{I}_{y0}&=\cos{\theta_\alpha}\mathcal{I}^{\text{c}}_{y0}-\sin{\theta_\alpha}\mathcal{I}^{\text{c}}_{zy},\notag\\
\mathcal{I}_{yx}&=-\cos{\theta_\beta}\sin{\theta_\delta}\mathcal{I}^{\text{c}}_{0y}
-\sin{\theta_\beta}\mathcal{I}^{\text{c}}_{yx}+\cos{\theta_\beta}\cos{\theta_\delta}\mathcal{I}^{\text{c}}_{yz},\notag\\
\mathcal{I}_{yy}&=\cos{\theta_\alpha}\sin{\theta_\delta}\mathcal{I}^{\text{c}}_{0z}-\sin{\theta_\alpha}\sin{\theta_\delta}\mathcal{I}^{\text{c}}_{xx}
+\cos{\theta_\alpha}\cos{\theta_\delta}\mathcal{I}^{\text{c}}_{yy}
-\sin{\theta_\alpha}\cos{\theta_\delta}\mathcal{I}^{\text{c}}_{z0},\notag\\
\mathcal{I}_{yz}&=\sin{\theta_\beta}\sin{\theta_\delta}\mathcal{I}^{\text{c}}_{0y}
-\cos{\theta_\beta}\mathcal{I}^{\text{c}}_{yx}-\sin{\theta_\beta}\cos{\theta_\delta}\mathcal{I}^{\text{c}}_{yz},\notag\\ 
\mathcal{I}_{z0}&=\sin{\theta_\alpha}\sin{\theta_\delta}\mathcal{I}^{\text{c}}_{0z}+\cos{\theta_\alpha}\sin{\theta_\delta}\mathcal{I}^{\text{c}}_{xx}+\sin{\theta_\alpha}\cos{\theta_\delta}\mathcal{I}^{\text{c}}_{yy}+\cos{\theta_\alpha}\cos{\theta_\delta}\mathcal{I}^{\text{c}}_{z0},\notag\\
\mathcal{I}_{zx}&=-\sin{\theta_\beta}\sin{\theta_\delta}\mathcal{I}^{\text{c}}_{x0}-\sin{\theta_\beta}\cos{\theta_\delta}\mathcal{I}^{\text{c}}_{zx}+\cos{\theta_\beta}\mathcal{I}^{\text{c}}_{zz},\notag\\
\mathcal{I}_{zy}&=\sin{\theta_\alpha}\mathcal{I}^{\text{c}}_{y0}+\cos{\theta_\alpha}\mathcal{I}^{\text{c}}_{zy},\notag\\
\mathcal{I}_{zz}&=-\cos{\theta_\beta}\sin{\theta_\delta}\mathcal{I}^{\text{c}}_{x0}-\cos{\theta_\beta}\cos{\theta_\delta}\mathcal{I}^{\text{c}}_{zx}-\sin{\theta_\beta}\mathcal{I}^{\text{c}}_{zz}.
\label{cafisospinrelation1}
\end{align}
where we defined $\mathcal{I}_{a0}\equiv\mathcal{S}_{a},\mathcal{I}_{0a}\equiv\mathcal{P}_{a},
\mathcal{I}_{ab}\equiv\mathcal{R}_{ab}$ and 
\begin{align}
\mathcal{I}_{0x}^{\text{c}}&=\text{Re}\left[\eta^\dagger_1\eta_3+\eta^\dagger_4\eta_2
-\eta^\dagger_4\eta_1 
-\eta^\dagger_2\eta_3
\right],\quad
\mathcal{I}_{0y}^{\text{c}}=\text{Im}\left[\eta^\dagger_1\eta_3+\eta^\dagger_4\eta_2
-\eta^\dagger_4\eta_1
-\eta^\dagger_2\eta_3
\right],\notag\\
\mathcal{I}_{0z}^{\text{c}}&=|\eta_4|^2-|\eta_3|^2,\notag\\
\mathcal{I}_{x0}^{\text{c}}&=\text{Re}[\eta_1+\eta_2],\quad
\mathcal{I}_{xx}^{\text{c}}=\text{Re}[\eta_3+\eta_4],\quad
\mathcal{I}_{xy}^{\text{c}}=\text{Im}[\eta_3-\eta_4],\quad
\mathcal{I}_{xz}^{\text{c}}=\text{Re}[\eta_1-\eta_2],
\notag\\
\mathcal{I}_{y0}^{\text{c}}&=\text{Im}[\eta_1+\eta_2],\quad
\mathcal{I}_{yx}^{\text{c}}=\text{Im}[\eta_3+\eta_4],\quad
\mathcal{I}_{yy}^{\text{c}}=-\text{Re}[\eta_3-\eta_4],\quad
\mathcal{I}_{yz}^{\text{c}}=\text{Im}[\eta_1-\eta_2],\notag\\
\mathcal{I}_{z0}^{\text{c}}&=1-\sum_{i=1}^4|\eta_i|^2,\quad
\mathcal{I}_{zx}^{\text{c}}=-\text{Re}\left[\eta^\dagger_1\eta_3+\eta^\dagger_4\eta_2
+\eta^\dagger_4\eta_1
+\eta^\dagger_2\eta_3
\right],\notag\\
\mathcal{I}_{zy}^{\text{c}}&=-\text{Im}\left[\eta^\dagger_1\eta_3+\eta^\dagger_4\eta_2
+\eta^\dagger_4\eta_1
+\eta^\dagger_2\eta_3
\right],\quad
\mathcal{I}_{zz}^{\text{c}}=|\eta_2|^2-|\eta_1|^2.
\label{cafisospinrelation2}
\end{align}
From \eqref{cafisospinrelation1}, \eqref{cafisospinrelation2}, and the equal-time commutation relations \eqref{nu2etaccr},
it can be verified that the SU(4) algebraic relation  
\begin{equation}
[\mathcal{I}_{\mu\nu}(\boldsymbol{x},t),\mathcal{I}_{\gamma\delta}(\boldsymbol{x},t)]=
i\delta(\boldsymbol{x}-\boldsymbol{y})f_{\mu\nu,\gamma\delta,\mu^\prime\nu^\prime}\mathcal{I}_{\mu^\prime\nu^\prime}(\boldsymbol{y},t),  
\end{equation}
is held. 
\end{document}